\documentclass{jfm}
\usepackage{graphicx}
\usepackage{bm}
\usepackage{newtxtext}
\usepackage{newtxmath}
\usepackage{natbib}
\usepackage{hyperref}
\usepackage{subcaption}
\usepackage{overpic}
\hypersetup{
    colorlinks = true,
    urlcolor   = blue,
    citecolor  = black,
}

\newcommand{\RomanNumeralCaps}[1]

\usepackage{amsmath,amssymb}
\hypersetup{
	bookmarksnumbered,
	linkcolor=blue,
	anchorcolor = blue,
}
\graphicspath{{./Fig/},{./}}
\usepackage{placeins} 

\usepackage{cleveref}
\crefname{figure}{figure}{figures}
\crefname{equation}{}{}
\crefformat{section}{\S#2#1#3}
\crefmultiformat{section}{\S\S~#2#1#3}{ and~#2#1#3}{, #2#1#3}{, and~#2#1#3}

\usepackage{siunitx}
\usepackage[table]{xcolor} 
\renewcommand{\absfooterflag}{}
\renewcommand{\pagelimitfooter}{}
\definecolor{mycol}{cmyk}{0,1,0,0}  

\title{Steady transport of active particles under continuous release in confined shear flow}

\author{Hanhan Zeng\aff{1} 
  \and Guoqian Chen\aff{1,2}\corresp{\email{gqchen@pku.edu.cn}}     }

\affiliation{\aff{1} School of Mechanics and Engineering Science, Peking University, Beijing, PR China
\aff{2}National Observation and Research Station of Coastal Ecological  Environments in Macao, Macau University of Science and  Technology, Macao SAR, PR China
}
\begin{document}
\maketitle

\begin{abstract}
Continuous release is a common and fundamental source condition in transport problems, yet theoretical treatments have largely been confined to downstream concentration development for passive solutes or fully developed cross-sectional distributions of active particles. We develop a spatial theory for the steady transport of active particles under continuous release in confined shear flows, resolving the concentration field from the inlet to the far field. Based on the Smoluchowski equation, we formulate a boundary-value problem with a point-source inlet-flux condition and boundary conditions in the cross-sectional space. Separation of the streamwise coordinate from the cross-sectional variables yields a non-self-adjoint generalized eigenvalue problem, whose spatial modes are superposed to construct the solution. The eigenproblem is solved within a Galerkin spectral framework; downstream-admissible modes are retained and their coefficients determined from the inlet flux through a weighted biorthogonal expansion. Excellent agreement with individual-based simulations validates the theory. Applying the theory to plane Poiseuille flow under convection-dominated conditions, we find pronounced position--orientation coherence for spherical particles in the early developing region: swimming and shear reorientation drive the population through successive angular stages, generating alternating off-centre and centreline accumulation regions downstream, while dispersion progressively weakens this coherence. Particle elongation enhances orientational alignment, producing an early three-peaked vertical profile and stronger, more persistent lateral accumulation downstream, while reducing the coherence of centreward migration and thereby suppressing secondary centreline accumulation. The streamwise marginal concentration varies non-monotonically, reflecting changes in the mean streamwise velocity, with its far-field limit given by the reciprocal of the asymptotic drift velocity.
\end{abstract}

\begin{keywords}
swimming/flying, active particle, transport
\end{keywords}


\section{Introduction}
Active particles are microscopic entities capable of autonomous motion by converting energy drawn from their surroundings or stored internally into directed movement \citep{schweitzer_brownian_2003}. 
They occur widely in natural and engineered systems, including swimming micro-organisms \citep{lauga_hydrodynamics_2009} and synthetic active particles \citep{jiang_active_2010}. In fluid environments, the transport of active particles differs fundamentally from that of passive tracers. In addition to ambient advection and stochastic fluctuations, active particles migrate relative to the surrounding fluid by self-propulsion \citep{bees_advances_2020}, whose direction is governed by orientational dynamics \citep{ezhilan_transport_2015} and influenced by interactions with boundaries \citep{kantsler_ciliary_2013}. Research on the locomotion of natural micro-organisms and the design and control of synthetic self-driven particles has developed rapidly, with applications in environmental science \citep{urso_smart_2023}, microfluidics \citep{katuri_artificial_2016}, biomedical engineering \citep{li_micronanorobots_2017} and microrobotics \citep{palagi_bioinspired_2018}. Accordingly, understanding the mechanisms governing active-particle motion and transport is of fundamental importance and has become an active frontier of current research \citep{bechinger_active_2016}.

Continuous release is a common and fundamental mode of material input in fluid transport. It occurs when material is supplied to a flow over an extended period, either at a localized position or through the inlet of a flow-through system. Examples range from dye and contaminant inputs in environmental flows \citep{ghisalberti_mass_2005} to drug release under physiological flow conditions \citep{abbasnezhad_analyzing_2023}. Related active-particle configurations include the continuous injection of swimming bacteria into microchannels \citep{didio_active_2025} and the transport of self-propelled Janus colloids in syringe-pump-driven capillary flows \citep{si_self-propelled_2020}. Unlike an instantaneous release, which concerns the temporal evolution of a finite particle population, continuous release under steady flow and release conditions can establish a steady-state concentration field extending downstream from the source. This distinction motivates the question of how the concentration field develops from the release region towards its far-field state.

Theoretical studies of continuous-release transport have so far focused primarily on passive solutes. \citet{gill_dispersion_1972} constructed the unsteady concentration field generated by time-dependent continuous sources through a superposition integral of instantaneous-source solutions. For a steady source, they obtained an approximate axial distribution of the cross-sectionally averaged concentration and treated the corresponding spatial boundary-value problem using an orthogonal-function expansion in the downstream region together with a boundary-layer solution near the source. \citet{rubol_vertical_2016} used a generalized integral transform to obtain a semi-analytical solution for the steady concentration field generated by a continuous solute source in a vegetated flow. \citet{guo_solute_2022} later developed an analytical solution for the complete steady concentration field and established its connection with the corresponding instantaneous-release problem; related steady continuous-source transport problems have also been investigated \citep{guo_dispersion_2023}.
These studies show that continuous release can produce a steady concentration field that develops spatially from the prescribed source distribution towards a fully developed far-field state.

Active-particle transport in fluid flows has also been the subject of extensive theoretical study. A well-developed line of work concerns particle dispersion and is generally based on an instantaneous-release setting \citep{brenner_theory_1982,frankel_foundations_1989,hill_taylor_2002,manela_generalized_2003}. In confined flows, the evolution of a finite particle population is characterized through moments of the streamwise position \citep{aris_dispersion_1956} and the resulting dispersion characteristics, such as the effective drift velocity and dispersivity \citep{bees_dispersion_2010,croze_gyrotactic_2017,jiang_dispersion_2019}. In the asymptotic dispersion regime, the zeroth streamwise moment, corresponding to the cross-sectional distribution with particle orientation retained, approaches a stationary state \citep{brenner_macrotransport_1993,jiang_dispersion_2019}, whereas the streamwise marginal concentration can be approximated by an effective macroscopic transport equation based on these dispersion characteristics \citep{bees_dispersion_2010,peng_upstream_2020}. These formulations therefore provide reduced-order descriptions of the transport process: they characterize the cross-sectional distribution and the streamwise marginal concentration separately, without directly resolving the streamwise development of the complete concentration field in position–orientation space. A related line of work has focused on steady cross-sectional distributions under assumptions that remove the explicit streamwise dependence \citep{bearon_spatial_2011,ezhilan_transport_2015,vennamneni_shear-induced_2020,Maretvadakethope_interplay_2023}. Collectively, these studies show how particle swimming and various aspects of orientational dynamics, including the effects of shear and boundaries, govern cross-stream migration and distribution. In a continuous-release problem, these distributions correspond to the cross-sectional structure of the fully developed far field. The complete downstream development of the steady active-particle concentration field under continuous release therefore cannot be determined from these existing theories alone.

To our knowledge, no previous theoretical study has addressed the steady transport of active particles continuously released into a confined shear flow. Nor has a spatial theoretical framework been developed to resolve the complete position--orientation concentration field from the source region to the fully developed far field. The orientation degree of freedom increases the dimensionality of the problem and couples particle transport in position space with dynamics in orientation space. Moreover, the local streamwise particle velocity depends on both cross-sectional position and orientation and may change sign across this space, allowing particle motion to be directed either downstream or upstream. These features introduce considerable complexity into the formulation and solution of the spatial transport problem. Compared with the fully developed far field, the developing concentration field may undergo a much richer and more complex evolution in both position and orientational structure along the downstream direction. Such downstream development has received little theoretical attention in previous studies of active-particle transport.

In this work, we develop a spatial theory for the steady transport of active particles under continuous release in a confined shear flow, resolving the concentration field from the inlet to the fully developed far field. Starting from the Smoluchowski equation, we establish the continuous-release formulation and pose the corresponding steady spatial boundary-value problem with a prescribed inlet flux and boundary conditions in the cross-sectional space. Separation of the streamwise coordinate from the cross-sectional variables leads to a non-self-adjoint generalized eigenvalue problem. The eigenvalue problem is solved within a Galerkin spectral framework, and the complete concentration field is constructed as a superposition of downstream-admissible spatial modes. Their coefficients are determined from the prescribed inlet flux through a weighted biorthogonal expansion. The theoretical results are validated against individual-based simulations. Applying the theory to plane Poiseuille flow under advection-dominated conditions, we characterize the downstream evolution through vertical concentration profiles, two-dimensional concentration fields, cross-sectional position–orientation distributions, and streamwise concentration distributions. The effects of self-propulsion and particle shape are also examined. This work provides an important basis for studying continuous-release transport in more complex active-particle systems and geometries.

The remainder of the paper is organized as follows. \Cref{sec 2} formulates the steady spatial boundary-value problem based on the Smoluchowski equation, including nondimensionalization, the point-source inlet-flux condition, the far-field condition and the boundary conditions in the cross-sectional space. Section~3 develops the spatial solution framework using separation of the streamwise and cross-sectional variables, a Galerkin spectral treatment of the resulting generalized eigenvalue problem, selection of downstream-admissible modes and a weighted biorthogonal expansion to reconstruct the complete concentration field. Section~4 first validates the theory against individual-based simulations and then examines the vertical profiles and two-dimensional, streamwise and cross-sectional concentration distributions, with particular attention to the effects of self-propulsion and particle shape. Section~5 summarizes the main conclusions and discusses possible future extensions. Appendix~A provides details of the individual-based numerical simulations.

\section{Formulation of transport problem}
\label{sec 2}
In this section, we formulate the steady transport of active particles continuously released into a confined shear flow as a spatial boundary-value problem, using a two-dimensional channel configuration as a representative setting. Based on the Smoluchowski equation, we specify the point-source inlet-flux condition, the far-field condition and the boundary conditions in the cross-sectional space that make the spatial problem well defined.
\subsection{Physical setting and governing equation}
\begin{figure}
	\includegraphics[scale=0.21]{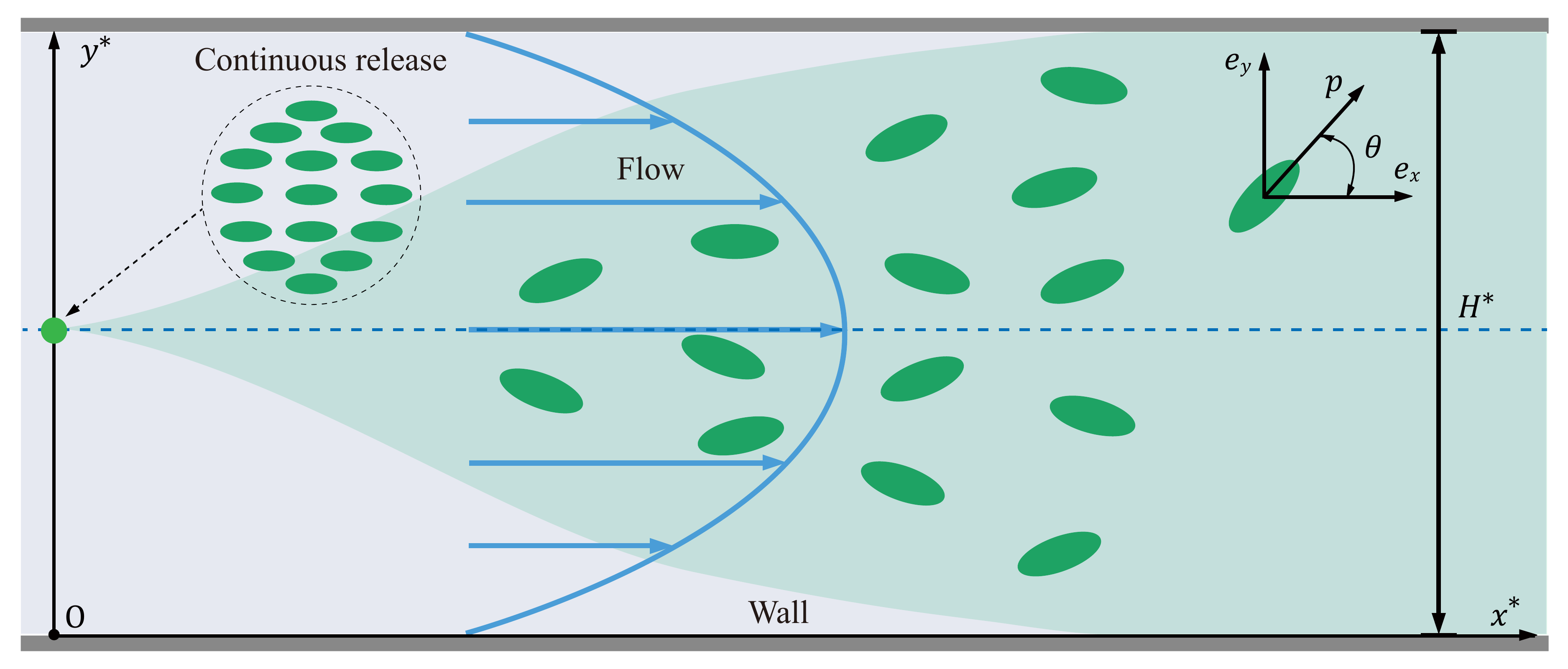}
	\caption{Schematic of the steady transport of active particles from a continuous release source in a plane Poiseuille flow.}
	\label{fig sketch}
\end{figure}
As shown in~\cref{fig sketch}, we consider active particles confined in a
two-dimensional channel bounded by two parallel walls separated by a distance
$H^{\ast}$. A Cartesian coordinate system $(x^{\ast},y^{\ast})$ is introduced,
where $x^{\ast}$ is the streamwise coordinate and $y^{\ast}$ is the vertical
coordinate, with corresponding unit vectors $(\bm{e}_x,\bm{e}_y)$. The imposed
background flow is a laminar unidirectional shear flow
$\bm{U}^{\ast}=U^{\ast}(y^{\ast})\bm{e}_x$, where $U^{\ast}(y^{\ast})$ denotes
the streamwise velocity profile. Throughout this paper, the superscript
$\ast$ denotes dimensional quantities, while dimensionless variables
introduced later are written without the superscript.
The particle orientation is described by the angle $\theta$ measured from the
positive $x^{\ast}$-axis, and the corresponding unit swimming-direction vector
is $\bm{p}=\cos\theta\,\bm{e}_x+\sin\theta\,\bm{e}_y$.
The variables $(y^{\ast},\theta)$ together define the cross-sectional space.
The particles are assumed to swim at a constant characteristic speed $V_s^{\ast}$, taken to represent the mean swimming speed of the suspension, and to undergo
translational and rotational diffusion with coefficients $D_t^{\ast}$ and
$D_r^{\ast}$, respectively. The suspension is assumed to be sufficiently
dilute that particle--particle interactions and feedback on the imposed flow
can be neglected~\citep{bees_advances_2020}.

Particles are continuously supplied at a constant total release rate
$Q^{\ast}$ from a localized source at $x^{\ast}=0$. Their distribution over the cross-sectional space $(y^{\ast},\theta)$ is prescribed by a normalized source distribution $S_0^{\ast}(y^{\ast},\theta)$.
We consider the most basic release configuration, namely a point-source
release. Specifically, the point source is located at the channel centreline
$y^{\ast}=H^{\ast}/2$, and all particles have the downstream orientation
$\theta=0$ at release. 
The corresponding normalized source distribution is
\begin{equation}
S_0^{\ast}(y^{\ast},\theta)
=\delta\left(y^{\ast}-\frac{H^{\ast}}{2}\right)
\delta_{2\pi}(\theta),
\label{eq dimensional point source}
\end{equation}
where $\delta$ is the Dirac delta function and $\delta_{2\pi}$ is its
$2\pi$-periodic counterpart.

We consider the resulting steady-state transport in the downstream half-space
$x^{\ast}>0$, with the source plane $x^{\ast}=0$ serving as the inlet boundary,
and denote the position--orientation concentration density by
$P^{\ast}(x^{\ast},y^{\ast},\theta)$.
The concentration density satisfies the
steady Smoluchowski equation in the combined position--orientation
space~\citep{doi_theory_1988},
\begin{equation}
\bm{\nabla}_R^{\ast}\bm{\cdot}
\left[
\left(\bm{U}^{\ast}+V_s^{\ast}\bm{p}\right)P^{\ast}
-D_t^{\ast}\bm{\nabla}_R^{\ast}P^{\ast}
\right]
+\bm{\nabla}_p\bm{\cdot}
\left(
\dot{\bm{p}}^{\ast}P^{\ast}
-D_r^{\ast}\bm{\nabla}_pP^{\ast}
\right)
=0.
\label{eq steady Smoluchowski}
\end{equation}
Here $\bm{\nabla}_R^{\ast}=\frac{\partial}{\partial x^{\ast}}\bm{e}_x+\frac{\partial}{\partial y^{\ast}}\bm{e}_y$ and $\bm{\nabla}_p=\frac{\partial}{\partial \theta}\bm{e}_{\theta}$ denote the position-space and orientation-space gradients, respectively. The rate of change of swimming direction is governed by Jeffery's equation \citep{jeffery_motion_1922,hinch_effect_1972}:
\begin{equation}
		\dot{\bm{p}}^{\ast}=
         \frac{1}{2} \bm{\omega}^{\ast} \times \bm{p}+\alpha_0 [\bm{p} \times (\bm{E}^{\ast}\bm{\cdot} \bm{p})] \times \bm{p},
\end{equation}
where 
$\bm{\omega}^{\ast}=\bm{\nabla}_R^{\ast}\times\bm{U}^{\ast}$ and $\bm{E}^{\ast}=\frac{1}{2}\left[\bm{\nabla}_R^{\ast}\bm{U}^{\ast}+(\bm{\nabla}_R^{\ast}\bm{U}^{\ast})^{\mathrm{T}}\right]$ are the vorticity and rate-of-strain tensor of the flow, respectively,
and $\alpha_0$ is the shape factor of the active particles ranging from $0$ (spherical) to $1$ (rod-like).

For the convection-dominated downstream transport considered here, we neglect
streamwise translational diffusion while retaining vertical translational
diffusion \citep{guo_solute_2022,guan_streamwise_2024}. The steady concentration field varies over a much longer length
scale in the streamwise direction than across the channel, and streamwise
translational diffusion is small compared with the effective longitudinal
dispersion generated by shear and swimming
\citep{jiang_dispersion_2019,peng_upstream_2020}. Its principal effect is
therefore expected to be confined to smoothing the localized inlet profile,
without materially altering the downstream spatial development considered
here. Under this approximation,~\cref{eq steady Smoluchowski} can be written in
expanded form as
\begin{align}
&\left[U^{\ast}(y^{\ast})+V_s^{\ast}\cos\theta\right]
\frac{\partial P^{\ast}}{\partial x^{\ast}}
+V_s^{\ast}\sin\theta\frac{\partial P^{\ast}}{\partial y^{\ast}}
-D_t^{\ast}\frac{\partial^2P^{\ast}}{\partial {y^{\ast}}^2}
\notag\\
&\quad
-\frac{\mathrm{d}U^{\ast}}{\mathrm{d}y^{\ast}}
\alpha_0\sin2\theta\,P^{\ast}
+\frac{1}{2}\frac{\mathrm{d}U^{\ast}}{\mathrm{d}y^{\ast}}
\left(-1+\alpha_0\cos2\theta\right)
\frac{\partial P^{\ast}}{\partial\theta}
-D_r^{\ast}\frac{\partial^2P^{\ast}}{\partial\theta^2}=0.
\label{eq conservation}
\end{align}

To express the continuous-release condition at the inlet, we define the streamwise particle velocity as
$\mathcal{V}^{\ast}(y^{\ast},\theta)
=U^{\ast}(y^{\ast})+V_s^{\ast}\cos\theta$.
With streamwise translational diffusion neglected,
$\mathcal{V}^{\ast}P^{\ast}$ represents the streamwise particle-number flux
density. Since $Q^{\ast}$ is the total release rate and
$S_0^{\ast}(y^{\ast},\theta)$ is the normalized source distribution,
$Q^{\ast}S_0^{\ast}(y^{\ast},\theta)$ is the corresponding release-rate
density in the cross-sectional space $(y^{\ast},\theta)$. Particle-number conservation at the inlet therefore gives
\begin{equation}
\mathcal{V}^{\ast}(y^{\ast},\theta)
P^{\ast}(0,y^{\ast},\theta)
=Q^{\ast}S_0^{\ast}(y^{\ast},\theta).
\label{eq dimensional inlet flux}
\end{equation}
We seek a downstream solution that approaches a fully developed state in the
far field, with no further variation in the streamwise direction. The
corresponding far-field condition is
\begin{equation}
\lim_{x^{\ast}\to\infty}
\frac{\partial P^{\ast}}{\partial x^{\ast}}
(x^{\ast},y^{\ast},\theta)=0.
\label{eq dimensional far field}
\end{equation}

\subsection{Dimensionless form}
The dimensionless parameters are defined as follows:
\begin{equation} \label{eq dimensionless variable}
\left. 
\begin{aligned} 
P&=\frac{P^{\ast} H^{\ast} U^{\ast}_m}{Q^{\ast}}, \quad
\mathit{Pe}_s = \frac{V_s^{\ast}}{D^{\ast}_r H^{\ast}}, \quad
\mathit{Pe}_f = \frac{U^{\ast}_m}{D^{\ast}_r H^{\ast}}, \quad
U = \frac{U^{\ast}}{U^{\ast}_m}, \\
x &= \frac{x^{\ast}}{Pe_fH^{\ast}}, \quad
y = \frac{y^{\ast}}{H^{\ast}}, \quad
D_t = \frac{D_t^{\ast}}{D_r^{\ast} {H^{\ast}}^2}, \quad
S_0=H^{\ast}S_0^{\ast}.
\end{aligned} 
\hspace{0em} \right\}
\end{equation}
Here $P(x,y,\theta)$ is the dimensionless concentration density;
$Pe_s$ is the swimming P\'eclet number comparing the swimming ability of the
microswimmers, or equivalently the strength of confinement;
$Pe_f$ is the flow P\'eclet number characterizing the imposed mean flow and
shear strength;
$D_t$ is the dimensionless vertical translational diffusivity; and $U_m^{\ast}$ is the mean flow velocity given as
\begin{equation}
	U_m^{\ast} \triangleq \frac{1}{H^{\ast}} \int_0^{H^{\ast}} U^{\ast}(y^{\ast}) \, \mathrm{d} y^{\ast}. 
\end{equation}
With the cross-sectional mean velocity as the velocity scale, the
dimensionless plane Poiseuille velocity profile is
\begin{equation}
U(y)=6y(1-y),\qquad 0\leq y\leq 1.
\label{eq Poiseuille profile}
\end{equation}

The dimensionless streamwise particle velocity is defined as
\begin{equation}
\mathcal{V}(y,\theta)
\triangleq \frac{\mathcal{V}^{\ast}(y^{\ast},\theta)}{U_m^{\ast}}
=U(y)+\frac{Pe_s}{Pe_f}\cos\theta .
\label{eq:streamwise_velocity_function}
\end{equation}
To make the streamwise spatial evolution of the concentration field explicit,
we collect all cross-sectional transport terms in an operator and write the
dimensionless probability conservation equation \cref{eq conservation} as
\begin{equation} 
\mathcal{V}\frac{\partial P}{\partial x} = \mathcal{L}P, 
\label{eq dimensionless}
\end{equation} 
where $\mathcal{L}(\cdot)$ denotes the transport operator acting in the
cross-sectional space $(y,\theta)$, defined by
\begin{align}
\mathcal{L}(\cdot) \, &\triangleq\, 
 D_t\frac{\partial^2 (\cdot)}{\partial y^2} + \frac{\partial^2 (\cdot)}{\partial \theta^2}  
- Pe_s \sin \theta \frac{\partial (\cdot)}{\partial y} \notag\\
&+Pe_f \frac{\mathrm{d} U}{\mathrm{d} y} \alpha_0 \sin 2\theta  (\cdot)-\frac{1}{2}Pe_f \frac{\mathrm{d} U}{\mathrm{d} y}(-1+ \alpha_0 \cos 2 \theta) \frac{\partial (\cdot)}{\partial \theta} \notag .
\label{eq operatorL}
\end{align}
The corresponding dimensionless inlet-flux and far-field conditions are
\begin{equation}
\left.
\begin{aligned}
\mathcal{V}(y,\theta)P(0,y,\theta)&=S_0(y,\theta),
\\
\lim_{x\to\infty}\frac{\partial P}{\partial x}(x,y,\theta)&=0 .
\end{aligned}
\right\}
\label{eq release condition}
\end{equation}
The transport problem in the downstream half-space $x\geq0$ is formulated with an inlet at which all particles enter, rather than leave, the domain; that is, particles present at $x=0$ have positive streamwise velocities. The prescribed point source satisfies this condition because particles are released at the channel centreline with a downstream orientation, where the streamwise particle velocity $\mathcal{V}(y,\theta)$ attains its maximum over the cross-sectional space. They therefore experience the strongest possible initial downstream transport and are rapidly carried away from the inlet. Moreover, under the convection-dominated conditions considered here (with the mean flow speed substantially exceeding the particle swimming speed), a negative streamwise velocity can arise only after particles have migrated sufficiently close to a wall and reoriented towards the upstream direction. By that stage, they have already been advected downstream, making subsequent recrossing of $x=0$ negligible. Accordingly, the inlet-flux condition in \cref{eq release condition} is imposed over the entire cross-sectional space.

\subsection{Boundary conditions in the cross-sectional space}
The inlet-flux condition has been specified in \cref{eq release condition}. We now impose the boundary conditions in the cross-sectional space to complete the formulation of the spatial transport problem.
The particle flux normal to the channel walls is the vertical flux,
$J_y=Pe_s\sin\theta\,P-D_t\partial P/\partial y$.
The channel walls are impermeable, so particles cannot leave the physical
domain through $y=0$ or $y=1$. Particle-number conservation within the
cross-sectional space therefore requires the zero-flux conditions \citep{bearon_spatial_2011,ezhilan_chaotic_2012}:
\begin{equation}
    \int_0^{2 \pi} J_y \,\mathrm{d}\theta=
    \int_0^{2\pi} \left(Pe_s\sin\theta\,P
    -D_t\frac{\partial P}{\partial y}\right)\mathrm{d}\theta=0,
    \qquad y=0,1.
    \label{eq zero flux}
\end{equation}
which represents that the total vertical particle flux at each wall is zero.

The integrated zero-flux condition alone does not determine how an active particle responds upon contact with a wall, and a more restrictive wall condition is therefore required \citep{bearon_trapping_2015,ezhilan_transport_2015}. In practice, the behaviour of active particles near walls is very complex and may involve adhesion, scattering and other processes \citep{cineros_reversal_2006,sipos_trapping_2015,zeng_sharp_2022}. A boundary condition in a continuum model therefore provides a coarse-grained
representation of the underlying particle--wall interactions \citep{Maretvadakethope_interplay_2023,zeng_dispersion_2025,fung_foundation_2025}.
Given that the present study focuses on steady, advection-dominated transport under continuous release and that the effects of detailed particle--wall interactions on the overall transport are expected to be relatively small under these conditions, we adopt the idealized specular-reflection condition, the most widely used wall condition in studies of confined active particles \citep{bearon_spatial_2011,jiang_dispersion_2019,Guan_migration_2024,wang_taylor_2025}.
Under this rule, a particle reaching a wall is reflected instantaneously, with the normal component of its swimming direction reversed and the tangential component unchanged \citep{volpe_simulation_2014}.

For the present vertical coordinate, specular reflection at the horizontal
walls gives
\begin{align}
	P(x,y,\theta)
    &=P(x,y,2\pi-\theta),
    \qquad y=0,1, \label{eq specular density}\\
	\frac{\partial P}{\partial y}(x,y,\theta)
    &=-\frac{\partial P}{\partial y}(x,y,2\pi-\theta),
    \qquad y=0,1.
    \label{eq specular derivative}
\end{align}
Periodic conditions are imposed in the orientational space:
\begin{align}
        P(x,y,0)&=P(x,y,2\pi),
        \label{eq orientation periodic density}\\
        \frac{\partial P}{\partial \theta}(x,y,0)
        &=\frac{\partial P}{\partial \theta}(x,y,2\pi).
        \label{eq orientation periodic derivative}
\end{align}

\section{Theoretical solution framework}
\label{sec_solution}
Here, we construct the complete steady concentration field through a superposition of separated spatial modes. Separation of the streamwise coordinate $x$ from the cross-sectional variables $(y,\theta)$ leads to a non-self-adjoint generalized eigenvalue problem. This problem is solved within a Galerkin spectral framework, after which the modes compatible with the far-field condition are retained to form the downstream-admissible expansion. The modal coefficients are determined from the prescribed inlet-flux condition through a velocity-weighted biorthogonal projection onto the corresponding left eigenfunctions. The resulting expansion reconstructs the full position--orientation concentration field and its marginal distributions throughout the downstream half-space.

\subsection{Separated spatial modes and the generalized eigenvalue problem}
With \cref{eq dimensionless} written as a spatial-evolution equation along the streamwise coordinate $x$, coupled to an operator acting in the cross-sectional space $(y,\theta)$, we construct the solution by following the idea of separation of variables. Specifically, we seek elementary separated spatial modes by separating the streamwise and cross-sectional dependences. A representative separated spatial mode is written as
\begin{equation}
P_k(x,y,\theta)=X_k(x)\Phi_k(y,\theta),
\label{eq:separated_mode}
\end{equation}
where $X_k(x)$ is the streamwise modal function and $\Phi_k(y,\theta)$ is the associated eigenfunction in the cross-sectional space.

Substitution of \cref{eq:separated_mode} into the governing equation \cref{eq dimensionless} gives
\begin{equation}
\mathcal{V}(y,\theta)\Phi_k(y,\theta)\frac{\mathrm{d}X_k}{\mathrm{d}x}
=X_k(x)\mathcal{L}\Phi_k(y,\theta).
\end{equation}
Separating the streamwise and cross-sectional dependences and introducing the separation constant $\mu_k$, we obtain
\begin{equation}
\left\{
\begin{aligned}
\mathcal{L}\Phi_k &= \mu_k\mathcal{V}\Phi_k,\\
\frac{\mathrm{d}X_k}{\mathrm{d}x} &= \mu_k X_k .
\end{aligned}
\right.
\label{eq:generalized_eigenvalue_problem}
\end{equation}
The first equation in \cref{eq:generalized_eigenvalue_problem} is a generalized eigenvalue problem in the cross-sectional space $(y,\theta)$, where the streamwise velocity $\mathcal{V}$ appears as the eigenvalue weight. The separation parameter $\mu_k$ is the spatial eigenvalue of the $k$th separated mode. The ordinary differential equation for $X_k$ gives
\begin{equation}
X_k(x)=\mathrm{e}^{\mu_k x},
\label{eq:x_eigenfunction}
\end{equation}
where the arbitrary multiplicative constant is included in the modal coefficient.

Since the governing equation and the cross-sectional boundary conditions are
linear, a formal solution may be represented as a superposition of separated
spatial modes:
\begin{equation}
P(x,y,\theta)=\sum_{k=1}^{\infty}a_kP_k(x,y,\theta)
=\sum_{k=1}^{\infty} a_k\mathrm{e}^{\mu_k x}\Phi_k(y,\theta),
\label{eq:separated_expansion}
\end{equation}
where $a_k$ is the modal coefficient associated with the $k$th separated mode.
The downstream condition will subsequently select the admissible modes, while
the inlet-flux condition determines the modal coefficients of the retained
modes.

The cross-sectional eigenfunctions $\Phi_k$ satisfy the boundary conditions inherited from $P$, namely
\begin{equation}
\left.
  \begin{aligned}
	\Phi_k(y,\theta) = \Phi_k(y,2\pi-\theta), \quad \mathrm{at} \; y = 0, 1,\\
	\frac{\partial \Phi_k(y,\theta)}{\partial y} = - \frac{\partial \Phi_k(y,2\pi-\theta)}{\partial y}, \quad \mathrm{at} \; y=0,1,\\
     \Phi_k(y,0) =  \Phi_k(y,2\pi), \\
		 \frac{\partial \Phi_k}{\partial \theta}(y,0) =  \frac{\partial \Phi_k}{\partial \theta}(y,2\pi) . 
\end{aligned}
\right \}
\label{eq_BC Phi}
\end{equation}
Together with these cross-sectional boundary conditions, the first equation in
\cref{eq:generalized_eigenvalue_problem} constitutes the non-self-adjoint
generalized eigenvalue problem governing the spatial eigenvalues $\mu_k$ and
the associated cross-sectional eigenfunctions $\Phi_k$.

\subsection{Galerkin spectral framework and downstream-admissible modes}
The generalized eigenvalue problem does not generally admit an analytical solution by classical techniques such as further separation of variables or standard Sturm--Liouville theory, owing to the coupled structure of the operator $\mathcal{L}$ in the cross-sectional variables and to the sign-changing nature of the velocity weight $\mathcal{V}$. We therefore solve it using a Galerkin spectral method \citep{doi_dynamics_1978,hill_taylor_2002}.

We construct a complete orthonormal spectral basis on the cross-sectional space from the eigenfunctions of the auxiliary Laplace operator $L_0 = \frac{\partial^2}{\partial y^2} + \frac{\partial^2}{\partial \theta^2}$ under the same boundary conditions. The non-self-adjoint operator $\mathcal{L}$ is then projected onto this basis. This construction incorporates the boundary conditions into the Galerkin space, so that each finite-dimensional approximation satisfies them automatically. Following related studies \citep{jiang_dispersion_2019,wang_gyrotactic_2022,zeng_tumbling_2025}, the basis functions satisfying the present specular-reflection condition \cref{eq_BC Phi} are taken as
\begin{align}
		\left.
		\begin{array}{c}
			\dfrac{1}{\sqrt{2\pi}},\quad
			\dfrac{1}{\sqrt{\pi}} \cos (n \pi y), \quad
			\dfrac{1}{\sqrt{\pi}}\cos (m \theta),\\[8pt]
			\sqrt{\dfrac{2}{\pi}}\cos (n \pi y)\cos (m \theta) ,\quad
			\sqrt{\dfrac{2}{\pi}} \sin (n \pi y)  \sin (m \theta),
		\end{array}
		\right \}
		\label{eq eigenfunctions reflective}
\end{align}
where $m = 1,2,\ldots$ and $n = 1,2,\ldots$, and $e_i$ denotes the $i$th basis function. The eigenfunctions $\Phi_k$ are expanded as
\begin{equation}
\Phi_k(y,\theta) = \sum_{i=1}^{\infty} r_{ik} e_i(y,\theta),
\label{eq eigenfunction expansion}
\end{equation}
where $\bm{r}_k=(r_{1k},r_{2k},\ldots)^{\mathrm{T}}$ is the coefficient vector of the $k$th eigenfunction with respect to the chosen basis.
For any two functions $g(y,\theta)$ and $h(y,\theta)$ on the cross-sectional space, we use the inner product
\begin{equation}
\langle g, h \rangle \triangleq \int_0^1 \int_{0}^{2\pi} \overline{g(y, \theta)} h(y, \theta) \, \mathrm{d}\theta \, \mathrm{d}y,
\label{inner product}
\end{equation}
where the overbar denotes complex conjugation.

Substituting \cref{eq eigenfunction expansion} into the generalized eigenvalue problem and projecting onto the basis functions gives the generalized matrix eigenvalue problem
\begin{equation}
\bm{L}\bm{r}_k=\mu_k\bm{B}\bm{r}_k,
\label{eq matrix generalized eigenvalue}
\end{equation}
where $L_{ij}$ and $B_{ij}$ denote the $(i,j)$ entries of the matrices $\bm{L}$ and $\bm{B}$, respectively, defined by
\begin{align}
L_{ij} &= \left\langle e_i, \mathcal{L}e_j \right\rangle, \\
B_{ij} &= \left\langle e_i, \mathcal{V}(y,\theta)e_j \right\rangle
= \frac{Pe_s}{Pe_f}\left\langle e_i,\cos\theta\,e_j\right\rangle
+\left\langle e_i,U(y)e_j\right\rangle,
\end{align}
for $i,j=1,2,\ldots$.
For the Poiseuille profile in \cref{eq Poiseuille profile}, $U(y)$ is quadratic and $U'(y)$ is linear, so the coefficient functions in $\mathcal{L}$ and $\mathcal{V}$ can be written as finite sums of separable terms involving low-order polynomials in $y$ and trigonometric functions in $\theta$. Together with the trigonometric basis functions, each separable operator term contributes a product of a one-dimensional polynomial--trigonometric integral in $y$ and a one-dimensional trigonometric integral in $\theta$. Hence the Galerkin matrix entries are finite sums of such products, which can be evaluated analytically and assembled into the Galerkin matrices \citep{jiang_dispersion_2020,zeng_dispersion_2026}.

After retaining a finite number of basis functions in the Galerkin expansion, \cref{eq matrix generalized eigenvalue} becomes a finite-dimensional generalized matrix eigenvalue problem. The matrix $\bm{L}$ is negative-semidefinite and nonsymmetric, reflecting the dissipative diffusion terms and the first-order swimming and shear-induced orientation terms in $\mathcal{L}$. The velocity-weight matrix $\bm{B}$ is symmetric but indefinite, because active self-propulsion makes $\mathcal{V}(y,\theta)$ change sign in the cross-sectional space; by contrast, the weight matrix in the passive-solute counterpart is positive definite \citep{guo_solute_2022}. This indefinite weight matrix precludes the usual positive-definite reductions, such as a Cholesky-based transformation to a standard eigenvalue problem. We therefore solve the generalized problem directly in its matrix-pencil form,
\begin{equation}
(\bm{L}-\mu\bm{B})\bm{r}=0.
\label{eq matrix pencil}
\end{equation}
The matrix pencil is solved using a QZ-type generalized Schur algorithm
\citep{demmel_generalized_1993,dongarra_chebyshev_1996}, which applies
orthogonal transformations directly to the matrix pair $(\bm{L},\bm{B})$ to
reduce it to a generalized Schur form. The finite non-zero generalized
eigenvalues $\mu_k$ are obtained from the resulting Schur pair, while the
neutral mode $\mu=0$ is determined separately from the null space of
$\bm{L}$, as discussed below.

For the downstream half-space considered here, each admissible separated mode
must be compatible with the far-field condition in \cref{eq release condition}.   Since each separated mode contains the streamwise factor $\exp(\mu_k x)$, this
condition requires $\partial_x[\exp(\mu_k x)\Phi_k]\to0$ as $x\to\infty$.  We therefore retain only the finite eigenvalues that are either exactly zero or have strictly negative real part:
\begin{equation}
\mathcal{K}_{\mathrm{r}}
=\{k:\mu_k\ \text{is finite and}\ 
[\mu_k=0\ \text{or}\ \operatorname{Re}(\mu_k)<0]\}.
\label{eq stable modes}
\end{equation}
Modes outside $\mathcal{K}_{\mathrm{r}}$ are discarded because they either grow
or remain oscillatory downstream. The neutral mode, denoted by $\Phi_1$, is determined separately by setting $\mu=0$ in the generalized eigenvalue problem. This gives the null-space problem
\begin{equation}
\mathcal{L}\Phi_1=0.
\label{eq neutral mode}
\end{equation}
In the Galerkin discretization, the corresponding coefficient vector
$\bm{r}_1$ satisfies
\begin{equation}
\bm{L}\bm{r}_1=\bm{0}.
\label{eq neutral mode matrix}
\end{equation}

We place the neutral mode first and relabel the retained decaying modes as
\begin{equation}
\lambda_1=0,\qquad
\lambda_n\triangleq\mu_{\kappa_n},\quad
\Phi_n\triangleq\Phi_{\kappa_n},\quad
\bm{r}_n\triangleq\bm{r}_{\kappa_n},
\qquad n=2,3,\ldots,
\label{eq stable mode relabelling}
\end{equation}
where $\{\kappa_n\}_{n\ge2}$ enumerates the indices in
$\mathcal{K}_{\mathrm{r}}$ associated with the decaying modes, ordered such that
\begin{equation}
\operatorname{Re}(\lambda_1)=0>
\operatorname{Re}(\lambda_2)\ge
\operatorname{Re}(\lambda_3)\ge\cdots .
\end{equation}

Accordingly, the formal expansion in \cref{eq:separated_expansion} reduces to
the downstream-admissible modal expansion
\begin{equation}
P(x,y,\theta)=\sum_{n=1}^{\infty}c_n\mathrm{e}^{\lambda_n x}\Phi_n(y,\theta)
=c_1\Phi_1(y,\theta)+\sum_{n=2}^{\infty}
c_n\mathrm{e}^{\lambda_n x}\Phi_n(y,\theta),
\label{eq retained mode expansion}
\end{equation}
where $c_n$ is the modal coefficient associated with the $n$th retained mode.

Since $\operatorname{Re}(\lambda_n)<0$ for $n\ge2$, all the decaying modes
vanish in the far field, and hence
\begin{equation}
P(x,y,\theta)\longrightarrow c_1\Phi_1(y,\theta)
\qquad\text{as}\qquad x\to+\infty .
\label{eq far field neutral mode}
\end{equation}
Thus, the neutral eigenfunction $\Phi_1$ determines the cross-sectional shape
of the fully developed far-field state. Up to normalization, this shape is the
same as the long-time asymptotic distribution of the zeroth streamwise moment
in the corresponding instantaneous-release problem.

\subsection{Weighted biorthogonal method}
Having constructed the downstream-admissible modal expansion, we now determine the modal coefficients $c_n$ by imposing the inlet-flux condition on the superposition of the retained modes. The cross-sectional eigenfunctions $\Phi_n$ appearing in this expansion are the right eigenfunctions of the generalized eigenvalue problem. Since this problem is non-self-adjoint, the right eigenfunctions are not generally orthogonal with respect to the standard inner product defined in \cref{inner product}, and the modal coefficients cannot be determined by a standard orthogonal projection. We therefore employ a biorthogonal expansion involving these right eigenfunctions and the corresponding left, or adjoint, eigenfunctions $\Psi_n$~\citep{strand_computation_1987,brezinski_biorthogonality_1992}. Because $\mathcal{V}$ serves as the eigenvalue weight in the generalized eigenvalue problem, the corresponding left--right biorthogonality relation is $\mathcal{V}$-weighted. The same weight also appears in the prescribed inlet flux, allowing the modal coefficients to be determined from the inlet-flux condition through the associated weighted biorthogonal projection.

With respect to the inner product defined in \cref{inner product}, integration by parts gives the formal adjoint differential operator
\begin{align}
\mathcal{L}^{\dagger}(\cdot) \, &\triangleq\,
D_t\frac{\partial^2(\cdot)}{\partial y^2}
+\frac{\partial^2(\cdot)}{\partial \theta^2}
\notag
+Pe_s\sin\theta\frac{\partial(\cdot)}{\partial y}
+\frac{1}{2}Pe_f\frac{\mathrm{d}U}{\mathrm{d}y}
\left(-1+\alpha_0\cos 2\theta\right)
\frac{\partial(\cdot)}{\partial \theta}.
\label{eq adjoint operatorL}
\end{align}
For each retained right eigenpair $(\lambda_n,\Phi_n)$, the corresponding left, or adjoint, eigenfunction $\Psi_n$ satisfies
\begin{equation}
\mathcal{L}^{\dagger}\Psi_n
=\overline{\lambda_n}\mathcal{V}\Psi_n,
\qquad n=1,2,\ldots .
\label{eq adjoint generalized eigenvalue}
\end{equation}
For the present specular-reflection and periodic conditions, the boundary terms in the adjoint identity vanish, and $\Psi_n$ satisfies the same boundary conditions as $\Phi_n$ in \cref{eq_BC Phi}.

Taking the inner products of the right eigenvalue problem with $\Psi_m$ and the adjoint eigenvalue problem for $\Psi_m$ with $\Phi_n$, and using the adjoint relation between $\mathcal{L}$ and $\mathcal{L}^{\dagger}$, gives
\begin{align}
\lambda_n\left\langle\Psi_m,\mathcal{V}\Phi_n\right\rangle
=\left\langle\Psi_m,\mathcal{L}\Phi_n\right\rangle
=\left\langle\mathcal{L}^{\dagger}\Psi_m,\Phi_n\right\rangle
=
\left\langle\overline{\lambda_m}\mathcal{V}\Psi_m,\Phi_n\right\rangle
=\lambda_m\left\langle\Psi_m,\mathcal{V}\Phi_n\right\rangle .
\label{eq weighted orthogonality derivation}
\end{align}
Hence
\begin{equation}
\left(\lambda_n-\lambda_m\right)
\left\langle\Psi_m,\mathcal{V}\Phi_n\right\rangle=0.
\label{eq weighted orthogonality identity}
\end{equation}
For $\lambda_n\ne\lambda_m$, this identity gives
$\langle\Psi_m,\mathcal{V}\Phi_n\rangle=0$. Thus, left and right eigenfunctions associated with distinct eigenvalues are biorthogonal under the $\mathcal{V}$-weighted pairing. Then we choose the left--right normalization such that
\begin{equation}
\left\langle\Psi_m,\mathcal{V}\Phi_n\right\rangle
=\delta_{mn},
\label{eq weighted biorthogonality}
\end{equation}
where $\delta_{mn}$ is the Kronecker delta.

The left eigenfunctions are expanded in the same spectral basis as the right eigenfunctions:
\begin{equation}
\Psi_n(y,\theta)=\sum_{j=1}^{\infty}\ell_{jn}e_j(y,\theta).
\label{eq adjoint eigenfunction expansion}
\end{equation}
Here $\bm{\ell}_n=(\ell_{1n},\ell_{2n},\ldots)^{\mathrm{T}}$ denotes the coefficient vector of $\Psi_n$ in the spectral basis.

Galerkin projection of the adjoint generalized eigenvalue problem gives the corresponding left matrix eigenvalue problem
\begin{equation}
\bm{L}^{\mathrm{H}}\bm{\ell}_n
=\overline{\lambda_n}\bm{B}^{\mathrm{H}}\bm{\ell}_n,
\label{eq left matrix eigenproblem}
\end{equation}
where the superscript $\mathrm{H}$ denotes the conjugate transpose. Taking the conjugate transpose of this equation gives the equivalent left-eigenvector form
\begin{equation}
\bm{\ell}_n^{\mathrm{H}}\bm{L}
=\lambda_n\bm{\ell}_n^{\mathrm{H}}\bm{B}.
\label{eq left matrix eigenproblem row}
\end{equation}
The weighted biorthogonality condition in \cref{eq weighted biorthogonality} is therefore expressed in the Galerkin coefficient space as
\begin{align}
\left\langle \Psi_m,\mathcal{V}\Phi_n\right\rangle
&=
\sum_{i,j}\overline{\ell_{im}}\,r_{jn}
\left\langle e_i,\mathcal{V}e_j\right\rangle
=
\bm{\ell}_m^{\mathrm{H}}\bm{B}\bm{r}_n
=
\delta_{mn}.
\label{eq matrix weighted biorthogonality}
\end{align}
Collecting the right and left eigenvectors into
$\bm{R}=(\bm{r}_1,\bm{r}_2,\ldots)$ and
$\bm{W}=(\bm{\ell}_1,\bm{\ell}_2,\ldots)$, this condition becomes
\begin{equation}
\bm{W}^{\mathrm{H}}\bm{B}\bm{R}=\bm{I}.
\label{eq matrix weighted biorthogonality compact}
\end{equation}

At the inlet plane, the downstream-admissible modal expansion becomes
\begin{equation}
P(0,y,\theta)=\sum_{n=1}^{\infty}c_n\Phi_n(y,\theta).
\label{eq biorthogonal expansion}
\end{equation}
Taking the $\mathcal{V}$-weighted pairing of \cref{eq biorthogonal expansion} with $\Psi_m$ and using \cref{eq weighted biorthogonality} gives
\begin{equation}
c_m=\left\langle\Psi_m,\mathcal{V}(y,\theta)P(0,y,\theta)\right\rangle,
\qquad m=1,2,\ldots .
\label{eq expansion coefficients preliminary}
\end{equation}
Equation~\eqref{eq biorthogonal expansion} represents the inlet-plane trace of the modal solution; the inlet data are supplied by the inlet-flux condition in \cref{eq release condition}. Since this condition is imposed over the entire cross-sectional space, it can be written as $\mathcal{V}(y,\theta)P(0,y,\theta)=S_0(y,\theta)$. Substitution into \cref{eq expansion coefficients preliminary} then gives
\begin{equation}
c_m=\left\langle\Psi_m,S_0(y,\theta)\right\rangle,
\qquad m=1,2,\ldots .
\label{eq expansion coefficients}
\end{equation}

For the dimensionless point source specified in \cref{eq dimensional point source}, \cref{eq expansion coefficients} gives
\begin{equation}
c_m=\left\langle\Psi_m,\delta\left(y-\frac{1}{2}\right)\delta_{2\pi}(\theta)\right\rangle=\overline{\Psi_m\left(\frac{1}{2},0\right)},
\qquad m=1,2,\ldots .
\label{eq point source modal coefficients}
\end{equation}
For its Galerkin representation, the point-source inlet flux is expanded in the chosen spectral basis, in the distributional sense, as
\begin{equation}
S_0(y,\theta)=\sum_{j=1}^{\infty}g_{0j}e_j(y,\theta),
\qquad
g_{0j}=\left\langle e_j,S_0\right\rangle
=\overline{e_j\left(\frac{1}{2},0\right)}
=e_j\left(\frac{1}{2},0\right),
\label{eq inlet flux expansion}
\end{equation}
where the last equality follows because the spectral basis functions are real-valued. Introducing the coefficient vectors
$\bm{g}_0=(g_{01},g_{02},\ldots)^{\mathrm{T}}$ and
$\bm{c}=(c_1,c_2,\ldots)^{\mathrm{T}}$, the Galerkin coefficient-space form of \cref{eq expansion coefficients} is
\begin{equation}
\bm{c}=\bm{W}^{\mathrm{H}}\bm{g}_0.
\label{eq matrix expansion coefficients}
\end{equation}

The governing equation and the cross-sectional boundary conditions imply conservation of the net streamwise particle flux through each cross-section, while the source normalization fixes its conserved value. Define the net particle flux through a cross-section at position $x$ by
\begin{equation}
\mathcal{F}(x)\triangleq
\int_0^1\int_0^{2\pi}
\mathcal{V}(y,\theta)P(x,y,\theta)\,
\mathrm{d}\theta\mathrm{d}y .
\label{eq streamwise particle flux}
\end{equation}
Integrating the governing equation\cref{eq dimensionless} over the cross-sectional space and using the conservative form of $\mathcal{L}$ gives
\begin{equation}
\frac{\mathrm{d}\mathcal{F}}{\mathrm{d}x}
=-\left[
\int_0^{2\pi}J_y\,
\mathrm{d}\theta
\right]_{y=0}^{y=1}
=0 .
\label{eq streamwise flux conservation}
\end{equation}
Here the boundary terms in orientation space vanish by periodicity, while the wall contributions vanish by the zero-normal-flux condition in \cref{eq zero flux}. Hence $\mathcal{F}(x)=\mathcal{F}(0)$ throughout the downstream domain. Using the inlet-flux condition $\mathcal{V}(y,\theta)P(0,y,\theta)=S_0(y,\theta)$, imposed over the entire cross-sectional space in \cref{eq release condition}, we obtain
\begin{equation}
\mathcal{F}(x)=\mathcal{F}(0)
=\int_0^1\int_0^{2\pi}
S_0(y,\theta)\,
\mathrm{d}\theta\mathrm{d}y
=1 .
\label{eq conserved source flux}
\end{equation}
The last equality follows directly from the normalized point-source distribution. Thus, the total streamwise particle flux is the relevant conserved quantity for the transport problem under continuous release.

\subsection{Solution reconstruction and numerical implementation}
Combining the retained downstream-admissible modes with the modal coefficients obtained from the weighted biorthogonal projection gives the steady concentration field in the downstream half-space:
\begin{equation}
P(x,y,\theta)=\sum_{n=1}^{\infty}
c_n\mathrm{e}^{\lambda_n x}\Phi_n(y,\theta).
\label{eq stable solution}
\end{equation}
Although the non-self-adjoint generalized eigenvalue problem may yield complex eigenpairs and hence complex modal coefficients, the governing operator, cross-sectional boundary conditions and prescribed inlet flux are all real. Consequently, the retained non-real eigenpairs can be chosen as complex-conjugate pairs $(\lambda_n,\Phi_n)$ and $(\overline{\lambda_n},\overline{\Phi_n})$, while the weighted projection yields the corresponding conjugate coefficients $c_n$ and $\overline{c_n}$. Each conjugate pair therefore makes a real contribution to \cref{eq stable solution}, ensuring that the reconstructed concentration field is real-valued.

To characterize the downstream transport, we define the orientation-integrated concentration $C(x,y)$ and the streamwise marginal concentration $C_x(x)$ as
\begin{align}
    C(x,y) &\triangleq \int_0^{2\pi} P(x,y,\theta)\,\mathrm{d}\theta, \\
    C_x(x) &\triangleq \int_0^1\int_0^{2\pi}P(x,y,\theta)\,\mathrm{d}\theta\mathrm{d}y
    = \int_0^1 C(x,y)\,\mathrm{d}y.
    \label{eq:marginal_concentration_definitions}
\end{align}
Here $C(x,y)$ represents the two-dimensional concentration distribution in physical space and, at each streamwise position, gives the corresponding vertical concentration profile.

To ensure numerical convergence at a reasonable computational cost, the basis functions in \cref{eq eigenfunctions reflective} are truncated with $N_y=120$ and $N_\theta=20$, yielding a Galerkin space of
$1+N_y+N_\theta+2N_yN_\theta=4941$ basis functions. For the numerical evaluation of the downstream-admissible expansion in \cref{eq stable solution}, approximately $10^3$ decaying modes with the largest real parts, in addition to the neutral mode, are retained for most parameter sets; the number of retained decaying modes is increased to 1400 when finer resolution of the entrance region is required.

For independent validation, we use individual-based numerical simulations of the dimensionless Langevin equations corresponding to the governing equation \cref{eq dimensionless}. The particle trajectories are advanced using the Euler--Maruyama scheme, with a fixed ensemble initialized by an instantaneous release from the prescribed source. Rather than introducing new particles continuously, the steady continuous-release concentration field is reconstructed by integrating the resulting instantaneous-release density over time. This strategy avoids the continually increasing particle number and computational cost associated with direct continuous injection and is consistent with the superposition principle underlying the continuum formulation. The theoretical and numerical simulation results are compared in Section~4.1, while further implementation details are given in Appendix~A.

\section{Results and discussion}
\label{sec_results1}
Guided by the parameter ranges considered in previous studies of active-particle transport under strong-flow conditions \citep{ezhilan_transport_2015,barry_shear-induced_2015,nili_population_2017,jiang_dispersion_2019}, we consider $Pe_s=1$ and $Pe_f=10$. Unless otherwise stated, these values are used throughout this section to represent a shear-dominated regime. For greater generality, a relatively small vertical translational diffusion is retained, with $D_t=10^{-4}$ throughout this section. To examine the role of particle shape in the steady spatial development, we compare spherical particles with $\alpha_0=0$ and elongated ellipsoidal particles with $\alpha_0=0.9$. We characterize the downstream transport through vertical concentration profiles, two-dimensional concentration distribution, cross-sectional position--orientation distributions, and streamwise concentration distribution. These results describe the evolution of the initially localized inlet distribution through the developing region towards the fully developed far field.

\subsection{Validation against individual-based simulations}

\begin{figure}
\centering
\includegraphics[scale=0.8]{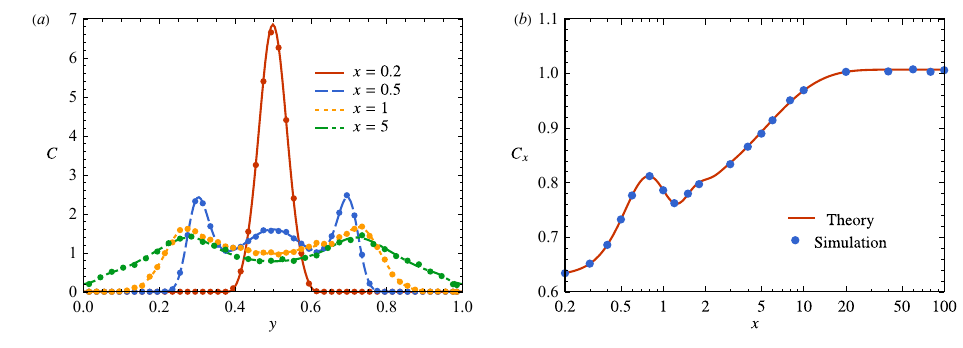}
\caption{Comparison between the theoretical model and individual-based numerical simulations for ellipsoidal active particles under continuous release. 
(\textit{a}) vertical concentration profiles $C$ at different streamwise positions $x$; 
(\textit{b}) streamwise concentration distribution $C_x$. 
Lines denote the theoretical results, and symbols denote the individual-based simulation results. 
Common parameters are $Pe_s=1$, $Pe_f=10$, $D_t=10^{-4}$ and $\alpha_0=0.9$.
}
\label{fig simulation comparison}
\end{figure}
We first validate the present theoretical model by comparison with individual-based numerical simulations. For this comparison, ellipsoidal active particles are used as a representative case. As shown in \cref{fig simulation comparison}, panel (a) presents the vertical concentration profiles $C$ at different streamwise positions $x$, while panel (b) shows the streamwise concentration distribution $C_x$. The theoretical and simulation results show excellent agreement in both quantities, including the locations and magnitudes of the concentration peaks and the non-monotonic variations along the downstream direction. This agreement validates the theoretical solution and demonstrates that the present model accurately captures the steady transport of active particles under continuous release.

\subsection{Vertical concentration profiles}
\label{subsec:vertical_concentration}
\subsubsection{Passive particles}

\begin{figure}
\centering
\includegraphics[scale=0.72]{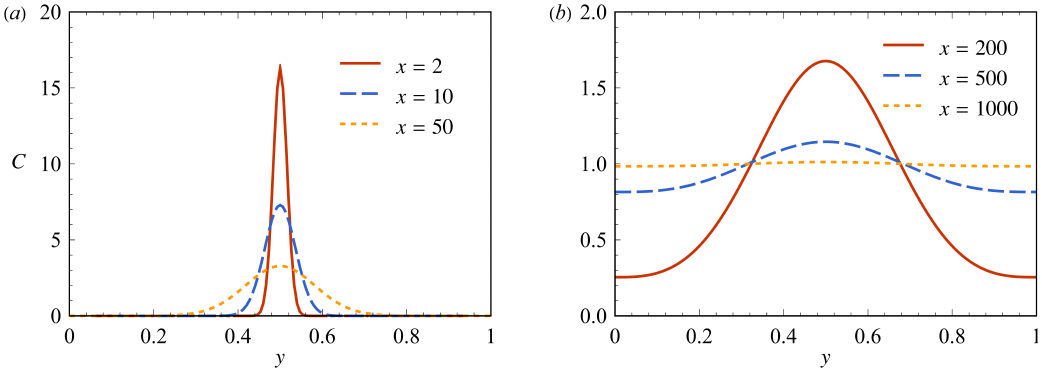}
\caption{Vertical concentration profiles $C(x,y)$ of passive particles at different streamwise position $x$ under continuous release.
(\textit{a}) $x=2$, $10$ and $50$;
(\textit{b}) $x=200$, $500$ and $1000$;
Common parameters are $Pe_s=0$, $Pe_f=10$, $D_t=10^{-4}$ and $\alpha_0=0$.
\label{fig Cy0}
}
\end{figure}
To isolate the effect of self-propulsion on the downstream concentration profiles, we first examine the simplest reference case of passive particles, corresponding to $Pe_s=0$.  In this case the streamwise
velocity no longer depends on orientation, $\mathcal{V}=U(y)$, and the dimensionless governing equation \cref{eq dimensionless} reduces to
\begin{equation}
U(y)\frac{\partial P}{\partial x}
=D_t\frac{\partial^2 P}{\partial y^2}
+\frac{\partial^2 P}{\partial \theta^2}
-\frac{\partial}{\partial \theta}
\left[
\frac{Pe_f}{2}\frac{\mathrm{d}U}{\mathrm{d}y}
\left(-1+\alpha_0\cos2\theta\right)P
\right].
\label{eq:passive_limit_density}
\end{equation}
Integrating \cref{eq:passive_limit_density} over $0\leq\theta\leq2\pi$ gives
\begin{equation}
U(y)\frac{\partial C}{\partial x}
=D_t\frac{\partial^2 C}{\partial y^2}.
\label{eq:passive_limit_Cy_equation}
\end{equation}
The rotational diffusion and Jeffery shear terms make no net contribution after integration because of periodicity in $\theta$.
Therefore, for passive particles with $Pe_s=0$, the governing equation in physical position space is the same as that for the classical solute-transport problem in a channel when streamwise diffusion is neglected.

\Cref{fig Cy0} shows the vertical concentration profiles of passive particles at different streamwise positions $x$ under continuous release. Near the inlet, the vertical concentration profile has a narrow centreline peak generated by the spreading of the point-source input: the concentration is strongly peaked at the channel centre and remains nearly zero near the walls. As $x$ increases, translational diffusion progressively broadens the vertical profile and reduces the centreline peak, while the distribution remains qualitatively single-peaked. At sufficiently large $x$, the concentration profile approaches the uniform fully developed far-field profile.
This follows from the governing equation: for passive particles, vertical redistribution is driven only by translational diffusion, so the vertical profile evolves over a relatively long streamwise length scale.

\subsubsection{Spherical active particles}
\begin{figure}
\centering
\includegraphics[scale=0.72]{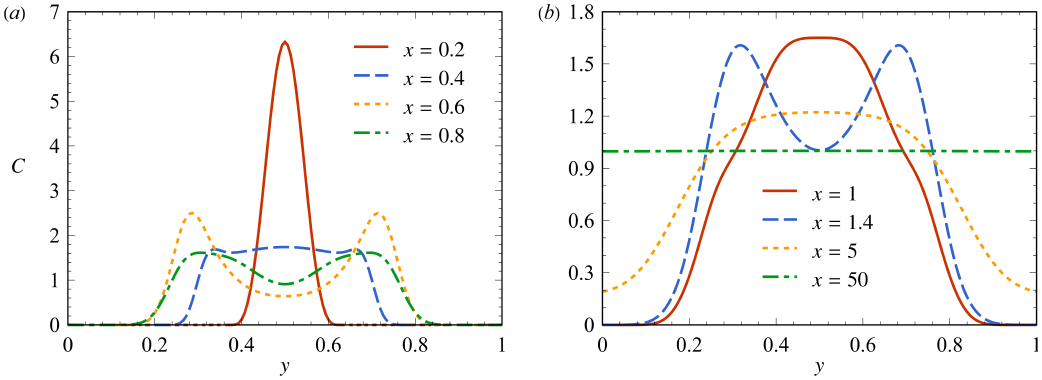}
\caption{Vertical concentration profiles $C$ of spherical active particles at different streamwise position $x$ under continuous release.
(\textit{a}) $x=0.2$, $0.4$, $0.6$, $0.8$;
(\textit{b}) $x=1$, $1.4$, $5$, $50$.
Common parameters are $Pe_s=1$, $Pe_f=10$, $D_t=10^{-4}$ and $\alpha_0=0$.
\label{fig Cy1}
}
\end{figure}
We next examine the vertical concentration profiles of spherical active particles at different streamwise positions $x$ under continuous release, as shown in \cref{fig Cy1}. Although the fully developed far-field vertical profile of spherical active particles is uniform, as in the passive-particle case, the developing profiles exhibit much richer and strongly non-monotonic evolution. This behaviour arises from active swimming and differs markedly from the solute spreading driven by advection and diffusion in previous continuous-release studies \citep{rubol_vertical_2016,guo_solute_2022}. In panel (a), the profile near the inlet initially takes the form of a narrow Gaussian-like distribution centred on the channel centreline. As $x$ increases, the centreline concentration decreases and the profile broadens, eventually splitting into two symmetric lateral peaks and forming an M-shaped distribution. Farther downstream in panel (b), the lateral peaks first weaken and merge by $x=1$, producing a broad centreline-peaked distribution. Two lateral peaks subsequently reappear at $x=1.4$, restoring the off-centre distribution, before merging again into a single broad centreline peak at $x=5$. The remaining vertical variation then gradually decays, and the profile eventually reaches the fully uniform state at $x=50$. These successive transitions between centreline and off-centre accumulation reflect a non-monotonic cross-stream redistribution of the particles, whose physical mechanism is examined more clearly using the two-dimensional concentration field in \cref{two_dimensional_concentration}.

\subsubsection{Ellipsoidal active particles}
\begin{figure}
\centering
\includegraphics[scale=0.72]{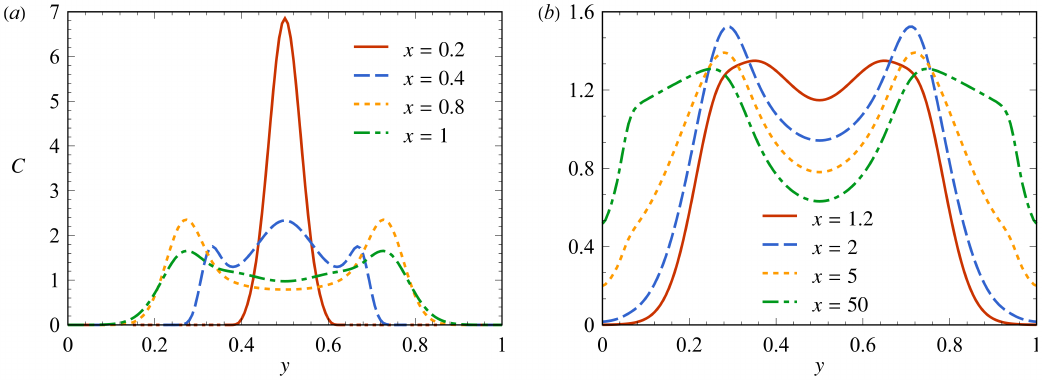}
\caption{Vertical concentration profiles $C$ of ellipsoidal active particles at different streamwise position $x$ under continuous release.
(\textit{a}) $x=0.2$, $0.4$, $0.8$, $1$;
(\textit{b}) $x=1.2$, $2$, $5$, $50$.
Common parameters are $Pe_s=1$, $Pe_f=10$, $D_t=10^{-4}$ and $\alpha_0=0.9$.
\label{fig Cy2}
}
\end{figure}

To investigate the effect of particle shape, we further present the vertical concentration profiles of ellipsoidal active particles at different streamwise positions $x$, as shown in \cref{fig Cy2}. In panel (a), the early profiles resemble those of spherical particles, with the narrow centreline peak becoming broader and lower downstream. At $x=0.4$, two lateral peaks emerge while the centreline peak remains pronounced, giving rise to an intermediate three-peaked distribution; farther downstream, the centreline peak gives way to a depression, yielding a double-peaked profile. In panel (b), the two lateral peaks initially move towards the centreline but, unlike those of spherical particles, do not merge to restore a centreline-peaked distribution. Still farther downstream, the double-peaked profile becomes more pronounced again by $x=2$ as the lateral peaks shift slightly towards the walls and the concentration near the centreline continues to decrease; thereafter, the near-wall concentration increases as the profile approaches a fully developed non-uniform state. Thus, shape-dependent shear reorientation alters both the developing profiles and their far-field distribution.

\subsection{Two-dimensional concentration distribution}
\label{two_dimensional_concentration}
\begin{figure}
	\centering
	\includegraphics[scale=0.6]{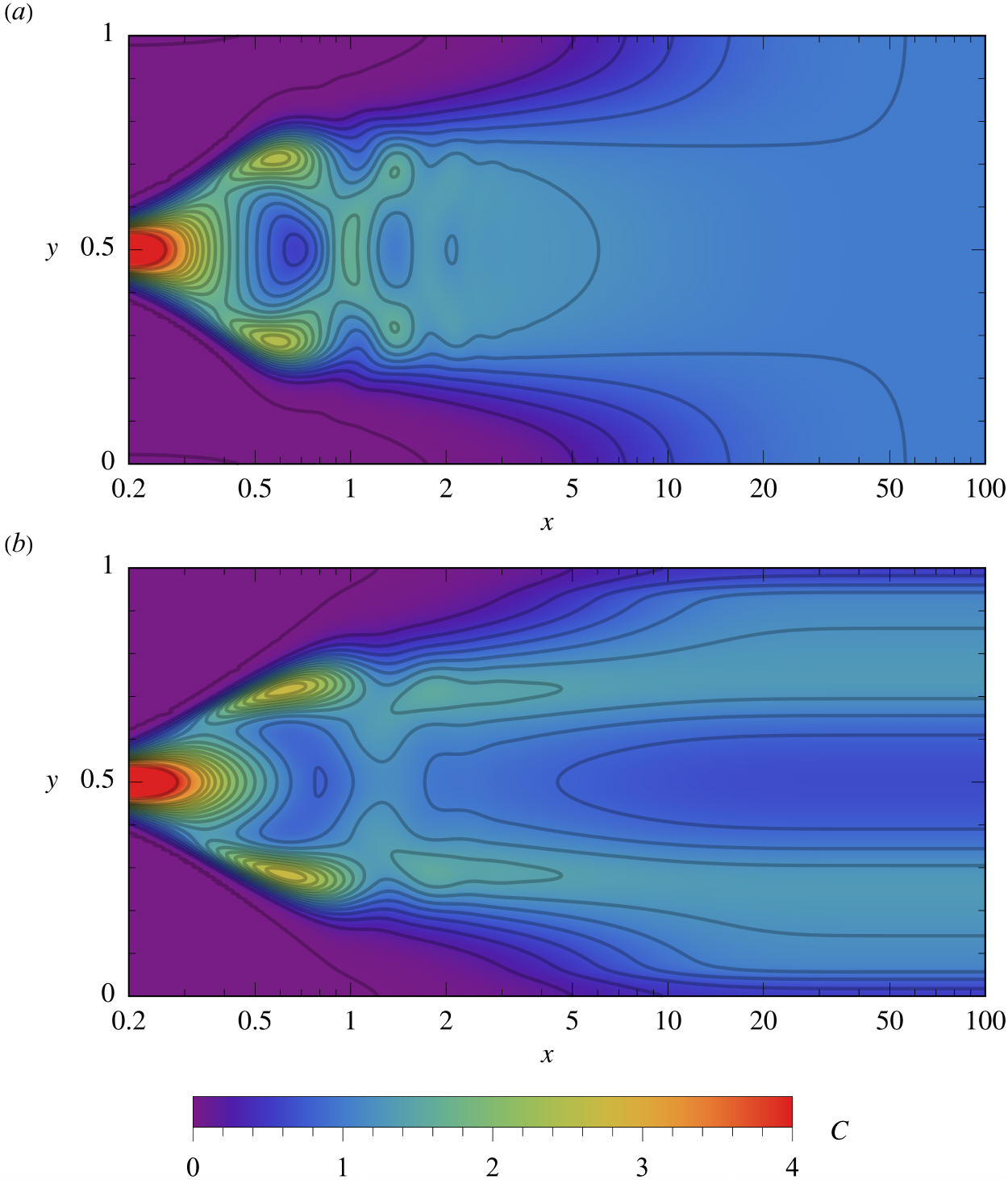}
	\caption{Two-dimensional concentration $C(x,y)$ of active particles under continuous release, with the streamwise coordinate $x$ plotted on a logarithmic scale.
    (\textit{a}) Spherical particles, $\alpha_0=0$; (\textit{b}) ellipsoidal particles, $\alpha_0=0.9$.
		Common parameters are $Pe_s=1$, $Pe_f=10$ and $D_t=10^{-4}$.
		\label{fig Cxy}
	}
\end{figure}

We further examine the two-dimensional concentration distributions of spherical and ellipsoidal active particles under continuous release through the contour plots of $C(x,y)$ in \cref{fig Cxy}, where the streamwise coordinate is shown on a logarithmic scale. For both particle shapes, the early developing region ($x\lesssim0.3$) is dominated by the narrow plume generated by the centreline point source, with the concentration centred at $y=0.5$. Although the particles are released in the downstream direction, rotational diffusion produces orientation deviations that generate vertical swimming components, allowing some particles to migrate away from the centreline. Once displaced from the centreline, the particles experience a local shear rate. Under the shear-dominated conditions considered here, the local shear tends to rotate the outward-moving particles from the downstream direction through the vertical direction and towards the upstream direction. Their outward swimming components therefore first increase and then decrease: above the centreline, the particles tend to rotate counterclockwise under the negative shear, whereas below the centreline, they tend to rotate clockwise under the positive shear. Coupled with downstream advection, this motion separates the centreline plume into two populations migrating towards opposite sides of the channel. As the particles approach the upstream direction, their wall-normal swimming components decrease towards zero, slowing their outward migration, while their net streamwise velocities are also reduced. The resulting increase in local residence time strengthens the two symmetric off-centre accumulation regions, which are most pronounced over approximately $0.5\lesssim x\lesssim1$.

For spherical particles, whose shear-induced angular velocity is independent of orientation, continued shear rotation carries a substantial fraction of the particles past the upstream direction and redirects them towards the centreline. The two off-centre accumulation regions consequently bend back towards the centreline and merge into a secondary centreline accumulation region near $x\simeq1$. This return and repeated cross-centreline migration are related to the single-particle swinging dynamics identified by \citet{zottl_nonlinear_2012,zottl_periodic_2013}, whose population-level manifestation under continuous release is the alternating formation of off-centre and centreline accumulation regions along the downstream direction. After the returning particles cross the centreline and enter the opposite half of the channel, the reversal of the local shear drives a weaker recurrence of outward and centreward migration, producing weaker off-centre accumulation regions followed by a still weaker third centreline accumulation near $x\simeq2$.  Dispersion broadens the particle distribution and gradually reduces the coherence of the collective cross-stream migration, so that no further distinct alternation between off-centre and centreline accumulation regions develops farther downstream. Instead, the plume spreads towards both walls and the concentration gradients gradually weaken. By $x=O(10)$, the localized accumulation regions have largely disappeared, and farther downstream, the concentration approaches the uniform fully developed distribution.

In contrast, orientation-dependent Jeffery rotation gives ellipsoidal particles a different downstream evolution. Near the downstream release orientation, their shear-induced angular velocity is small, so the inlet-centred plume decays more slowly and spans a larger streamwise scale than for spherical particles. Consequently, the centreline peak remains pronounced as the two lateral peaks develop, producing a three-peaked vertical concentration profile over approximately $0.4\lesssim x\lesssim0.6$. The Jeffery angular velocity is also small as the particles approach the upstream direction, increasing their residence near the lateral turning regions and thereby reinforcing the off-centre accumulation. This orientation-dependent slowing causes particles to return towards the centreline over a wider range of downstream positions, reducing the coherence of their centreward migration. They therefore do not converge to form a comparable secondary centreline accumulation. Instead, the two lateral accumulation regions extend downstream as broad bands, while the centreline concentration shows only a weak recovery before decreasing again. Despite this slower initial evolution, the absence of repeated off-centre--centreline redistribution makes the subsequent relaxation more direct than for spherical particles. As dispersion broadens the bands towards the walls, the concentration near the centreline continues to decrease while that near the walls increases, and the vertical distribution gradually approaches its fully developed profile.

\subsection{Cross-sectional position–orientation distributions}
\begin{figure}
\centering
\includegraphics[scale=0.72]{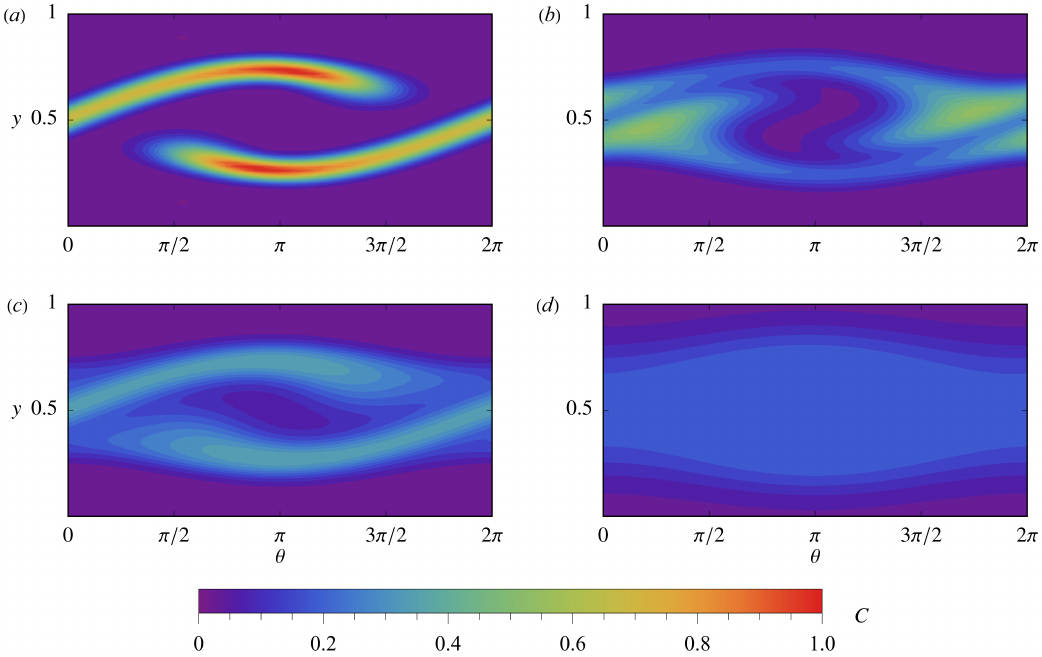}
\caption{Cross-sectional distribution $P(x,y,\theta)$ of spherical active particles at different streamwise position $x$ under continuous release.
(\textit{a}) $x=0.6$; (\textit{b}) $x=1$;
(\textit{c}) $x=1.4$; (\textit{d}) $x=5$.
Common parameters are $Pe_s=1$, $Pe_f=10$, $D_t=10^{-4}$ and $\alpha_0=0$.
\label{fig Cytheta1}
}
\end{figure}

\begin{figure}
\centering
\includegraphics[scale=0.72]{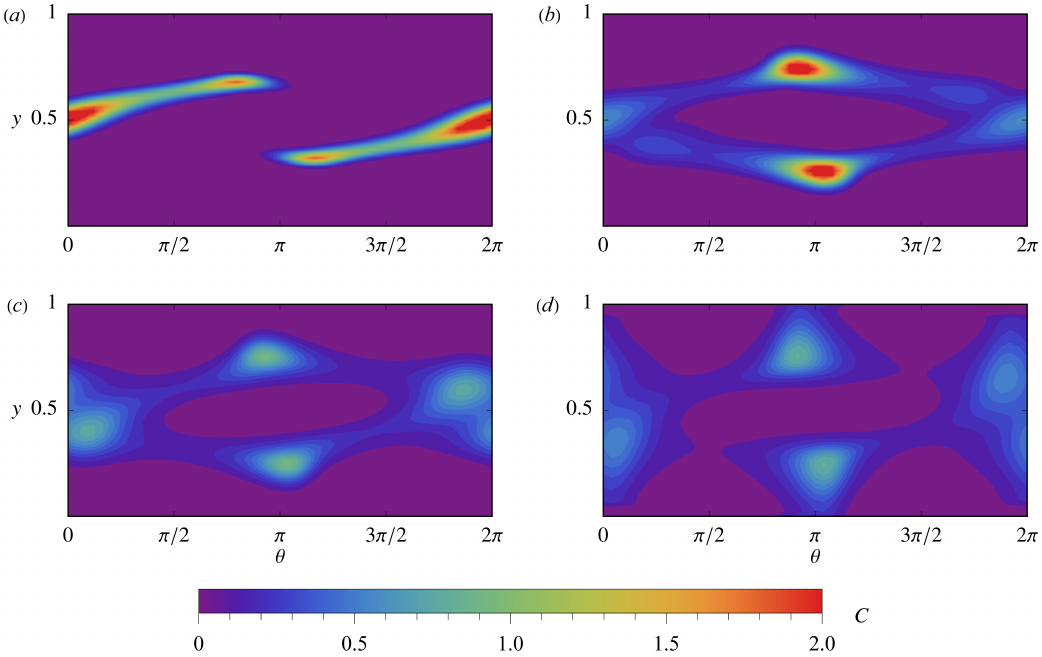}
\caption{Cross-sectional distribution $P(x,y,\theta)$ of ellipsoidal active particles at different streamwise position $x$ under continuous release.
(\textit{a}) $x=0.4$; (\textit{b}) $x=0.8$;
(\textit{c}) $x=1.2$; (\textit{d}) $x=5$.
Common parameters are $Pe_s=1$, $Pe_f=10$, $D_t=10^{-4}$ and $\alpha_0=0.9$.
\label{fig Cytheta2}
}
\end{figure}
\Cref{fig Cytheta1} shows the cross-sectional concentration distribution $P(x,y,\theta)$ of spherical active particles at different streamwise positions under continuous release. At $x=0.6$, the concentration is confined primarily to two narrow, continuous curved ridges, revealing pronounced position--orientation coherence, with particle orientation closely correlated with cross-stream position.
Such developing-region coherence has not, to our knowledge, been reported in
previous cross-sectional studies of active particles
\citep{jiang_transient_2021,Maretvadakethope_interplay_2023}.
These ridges correspond to the first off-centre accumulation, with their maxima near the upstream orientation $\theta=\pi$ marking the lateral turning of the shear-reoriented populations, after which continued rotation redirects them towards the centreline. At $x=1$, weaker accumulation occurs along the two branches approaching the centreline, where the particles are predominantly centreward-oriented and progressively realign with the downstream direction; a broader and weaker repetition of the outward-oriented branches appears at $x=1.4$. These distributions show that the same particle populations undergo a coherent angular progression: rotation from downstream to upstream alignment drives outward cross-stream migration, continued rotation towards downstream alignment drives centreward migration, and subsequent reorientation directs them outwards again, producing alternating off-centre and centreline accumulation regions downstream.
Dispersion progressively spreads the particles over different angular stages and damps this repeated redistribution; by $x=5$, the angular variation is weak although the vertical concentration remains non-uniform, showing that position--orientation coherence is lost before complete cross-stream homogenization.

\Cref{fig Cytheta2} shows the corresponding distributions for ellipsoidal active particles. Compared with the relatively simple orientational structure of spherical particles, ellipsoidal particles exhibit a more heterogeneous distribution involving several distinct angular populations. At $x=0.4$, slow rotation near flow alignment preserves a remnant centreline population in the downstream direction, while particles undergoing larger rotational-diffusive reorientation migrate laterally and are subsequently rotated by shear towards the upstream direction. The three-peaked vertical profile therefore results from the superposition of these two types of populations, which have different orientations and transport histories.
At $x=0.8$, a downstream-oriented component remains near $\theta=0$, whereas the pronounced lateral accumulations are concentrated near the upstream orientation $\theta=\pi$. At $x=1.2$, populations near the upstream and downstream directions coexist, indicating that upstream-oriented lateral particles remain while part of the population has already turned towards the centreline; this less coherent centreward migration suppresses the pronounced secondary centreline accumulation observed for spherical particles. From $x=1.2$ to $5$, the preferred orientations change little, and the downstream evolution consists mainly of progressive cross-stream broadening towards the walls. This wallward spreading is more extensive than for spherical particles, and the distribution gradually approaches a non-uniform fully developed state.

\subsection{Streamwise concentration distribution}
\subsubsection{Streamwise concentration profiles}
\label{subsec:streamwise_concentration}
\begin{figure}
\centering
\includegraphics[scale=0.9]{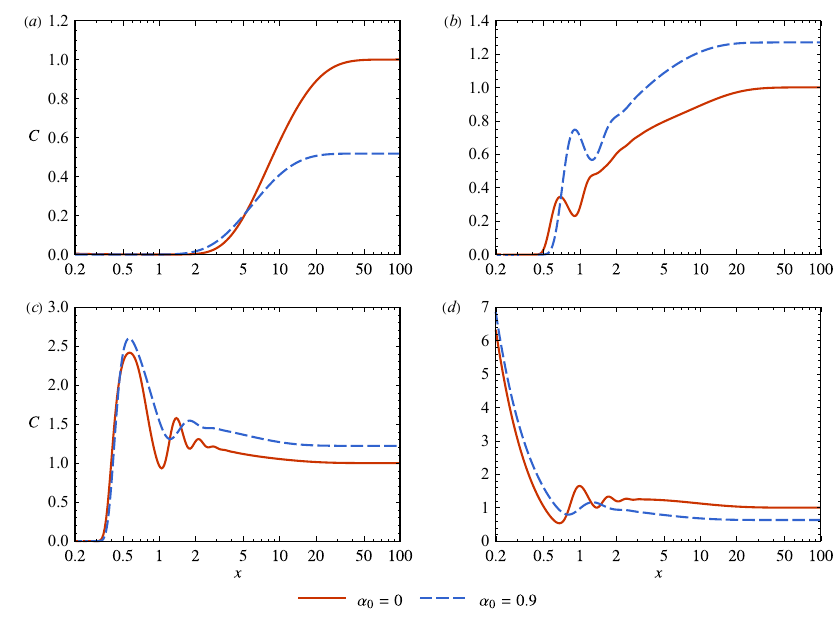}
\caption{Streamwise concentration profiles $C(x,y)$ of active particles at different vertical position $y$ under continuous release.
(\textit{a}) $y=0$; (\textit{b}) $y=0.2$;
(\textit{c}) $y=0.3$; (\textit{d}) $y=0.5$.
The solid and dashed curves represent spherical particles ($\alpha_0=0$) and ellipsoidal particles ($\alpha_0=0.9$), respectively.
Common parameters are $Pe_s=1$, $Pe_f=10$ and $D_t=10^{-4}$.
\label{fig Cx-y}
}
\end{figure}

\Cref{fig Cx-y} quantifies the downstream evolution of the two-dimensional fields in \cref{fig Cxy} through streamwise concentration profiles at different vertical positions. For $x\lesssim0.3$, the concentration variation remains mainly confined to the centreline region. Over $0.3\lesssim x\lesssim2$, redistribution occurs primarily between the centreline and the two off-centre regions. For spherical particles, the profiles at $y=0.5$ and $y=0.3$ each exhibit three successively smaller maxima, with their peak--trough variations occurring approximately out of phase. This pattern reflects repeated transitions between centreline and off-centre accumulation, while the decreasing amplitudes indicate that dispersion progressively weakens the coherence of the cross-stream redistribution. Only after these alternating variations have ceased, for $x\gtrsim2$, does the distribution begin to spread appreciably towards the walls, as indicated by the increase in wall concentration. Among the positions considered, only the wall concentration varies monotonically with $x$; all the interior profiles are non-monotonic owing to the successive passage and recurrence of the accumulation regions.

Compared with spherical particles, ellipsoidal particles show a slower early decay of the centreline concentration, so that the centreline peak persists farther downstream and the transition to a two-peaked profile is delayed. These stronger and broader off-centre bands spread to the wall region
earlier, as shown by the earlier rise of the wall profile. During the developing stage, the higher concentrations at $y=0.3$ and $y=0.2$ indicate stronger off-centre
accumulation, enhanced by the longer particle residence near the lateral turning
regions.  Despite their slower
early evolution, ellipsoidal particles do not undergo the repeated coherent
centreline--off-centre redistribution observed for spherical particles. Their
subsequent relaxation is therefore more direct, and the profiles approach the
fully developed M-shaped distribution over a shorter streamwise distance.

\subsubsection{Streamwise marginal concentration and far-field limit}
\begin{figure}
	\centering
	\includegraphics[scale=0.6]{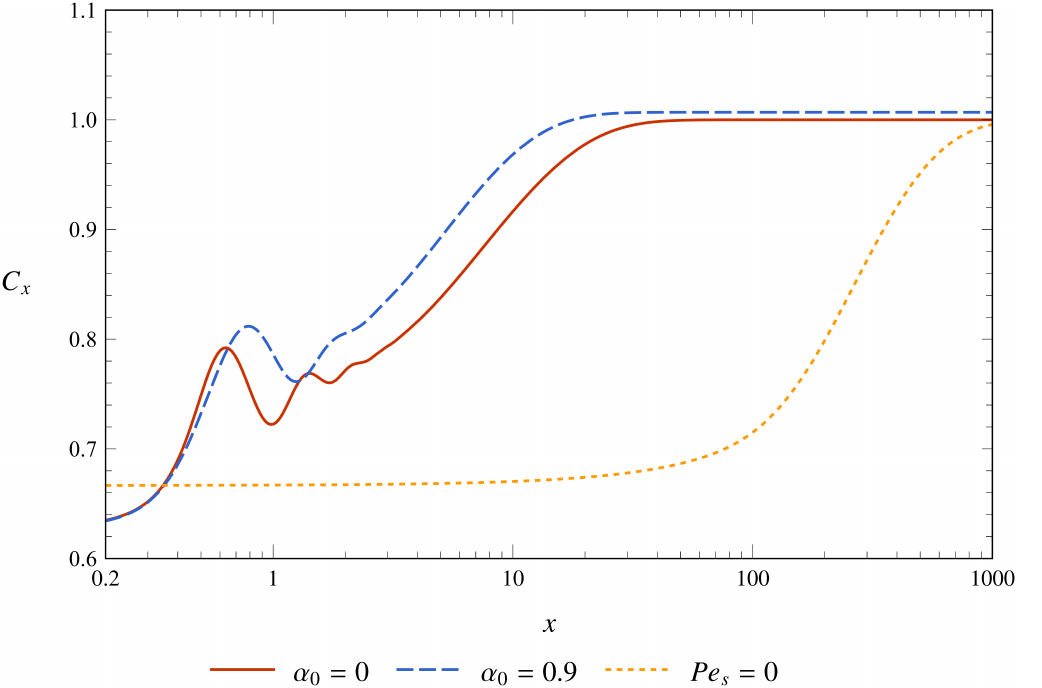}
	\caption{Streamwise marginal concentration $C_x(x)$ for spherical active particles ($\alpha_0=0$), ellipsoidal active particles ($\alpha_0=0.9$) and passive particles ($Pe_s=0$) under continuous release.
		For the active-particle cases, $Pe_s=1$; common parameters are $Pe_f=10$ and $D_t=10^{-4}$.
		\label{fig Cx}
	}
\end{figure}

\Cref{fig Cx} shows the downstream evolution of the streamwise marginal
concentration for active and passive particles. The concentration-weighted mean streamwise particle velocity is defined using the concentration distribution $P(x,y,\theta)/C_x(x)$ normalized over the cross-sectional space as
\begin{equation}
    U_C(x) \triangleq
    \int_0^1\int_0^{2\pi}
    \mathcal{V}(y,\theta)
    \frac{P(x,y,\theta)}{C_x(x)}
    \,\mathrm{d}\theta\,\mathrm{d}y .
    \label{eq local mean streamwise velocity}
\end{equation}
Using the conservation of the net streamwise flux, $\mathcal{F}(x)=1$, established in \cref{eq conserved source flux}, this becomes
\begin{equation}
    U_C(x)
    =\frac{1}{C_x(x)}
    \int_0^1\int_0^{2\pi}
    \mathcal{V}(y,\theta)P(x,y,\theta)
    \,\mathrm{d}\theta\,\mathrm{d}y
    =\frac{\mathcal{F}(x)}{C_x(x)}
    =\frac{1}{C_x(x)} .
    \label{eq mean velocity concentration relation}
\end{equation}
Thus, a larger $C_x$ corresponds to a lower concentration-weighted mean streamwise particle velocity over the cross-sectional space, resulting in a longer residence time per unit streamwise distance.

Both active-particle curves vary non-monotonically, first rising to a local
maximum, then decreasing to a shallow minimum and finally approaching their
far-field plateaus. The first maximum of the ellipsoidal-particle curve is
higher and occurs farther downstream than that of the spherical-particle
curve; its larger magnitude reflects stronger off-centre accumulation and
longer residence in states with lower streamwise velocities, whereas its
downstream shift is consistent with the slower early evolution of ellipsoidal
particles. The subsequent peak--trough variations of the spherical-particle
curve reflect repeated transitions between centreline and off-centre
accumulation; as this redistribution progressively loses coherence, the
variations weaken downstream. The ellipsoidal-particle curve instead
approaches its plateau more directly, becoming nearly constant by
$x\simeq20$, compared with $x\simeq30$ for spherical particles. In contrast,
the passive-particle concentration is initially higher owing to its lower
streamwise velocity, increases monotonically and approaches its plateau only
near $x=10^3$, because its wall-normal redistribution is driven solely by the
small translational diffusivity.

To determine the far-field plateau values approached by these curves, we
consider the neutral mode of the present spatial theory. Once all decaying
modes have vanished, \cref{eq far field neutral mode} gives
$P\to c_1\Phi_1$. Consequently, the normalized cross-sectional concentration
distribution in \cref{eq local mean streamwise velocity} tends to
\begin{equation}
    \rho_\infty(y,\theta)
    \triangleq
    \frac{\Phi_1(y,\theta)}
    {\int_0^1\int_0^{2\pi}
    \Phi_1(y,\theta)\,\mathrm{d}\theta\,\mathrm{d}y},
    \label{eq normalized far-field distribution}
\end{equation}
whose integral over the cross-sectional space is unity.
This normalized distribution coincides with the long-time asymptotic
distribution of the zeroth streamwise moment in the corresponding dispersion
problem initiated by a unit instantaneous release. The corresponding
concentration-weighted mean streamwise particle velocity, termed the drift
velocity in dispersion theory, follows from
\cref{eq local mean streamwise velocity} as
\begin{equation}
    U_d=\int_0^1\int_0^{2\pi}
    \mathcal{V}(y,\theta)\rho_\infty(y,\theta)
    \,\mathrm{d}\theta\,\mathrm{d}y.
    \label{eq drift velocity}
\end{equation}
Taking the far-field limit of
\cref{eq mean velocity concentration relation} then gives
\begin{equation}
    C_x^\infty
    \triangleq \lim_{x\to\infty}C_x(x)
    =\frac{1}{U_d},
    \qquad
    P_\infty(y,\theta)
    \triangleq \lim_{x\to\infty}P(x,y,\theta)
    =C_x^\infty\rho_\infty(y,\theta)
    =\frac{\rho_\infty(y,\theta)}{U_d}.
    \label{eq far-field concentration}
\end{equation}

The same far-field relation can also be derived from dispersion theory. Let
$C_I(x,t)$ denote the streamwise marginal concentration generated
by a unit instantaneous release at $x=0$. On sufficiently large temporal and
streamwise scales, its evolution is governed by the advection-diffusion equation
\begin{equation}
\frac{\partial C_I}{\partial t}
+U_d\frac{\partial C_I}{\partial x}
=D_T\frac{\partial^2C_I}{\partial x^2},
\end{equation}
where $D_T$ is the long-time asymptotic dispersion coefficient. For a unit impulse, the fundamental solution is
\begin{equation}
C_I(x,t)=\frac{1}{\sqrt{4\pi D_Tt}}
\exp\left[-\frac{(x-U_dt)^2}{4D_Tt}\right].
\end{equation}
For a source maintained at unit release rate, the steady streamwise marginal
concentration in the far field is given by the accumulated impulse response:
\begin{equation}
\begin{aligned}
C_x(x)
&\sim
\int_0^\infty C_I(x,t)\,\mathrm{d}t=
\int_0^\infty
\frac{1}{\sqrt{4\pi D_Tt}}
\exp\left[
-\frac{(x-U_dt)^2}{4D_Tt}
\right]\mathrm{d}t\\
&=
\frac{1}{U_d}
\exp\left(
\frac{U_dx-\lvert U_dx\rvert}{2D_T}
\right)
=\frac{1}{U_d},\qquad x>0,\quad U_d>0.
\end{aligned}
\end{equation}
This gives the same far-field relation as that derived directly from the
neutral spatial mode. It should be noted that (4.8) is valid only on large temporal and streamwise scales and cannot be used to describe the developing region.

The fully developed cross-sectional distributions of passive and spherical
active particles are both uniform; since angular averaging eliminates the
swimming contribution in the latter case, the mean-flow normalization yields
$U_d=1$ and hence $C_x^\infty=1$ for both. For the ellipsoidal active particles, the
non-uniform far-field distribution assigns greater weight to states with lower
streamwise velocities, resulting in $U_d<1$ and consequently
$C_x^\infty>1$. Their streamwise marginal concentration therefore approaches a
plateau slightly above unity.

\section{Concluding remarks}
\label{sec_conclusion}

We develop a spatial theory for the steady transport of active particles under continuous release in confined shear flows, resolving the concentration field from the inlet to the far field. Based on the Smoluchowski equation, the continuous-release problem is posed as a steady spatial boundary-value problem with a point-source inlet-flux condition, a far-field condition and boundary conditions in the cross-sectional space. Separation of the streamwise coordinate from the cross-sectional variables yields a non-self-adjoint generalized eigenvalue problem, and the solution is represented as a superposition of the resulting spatial modes. To handle the strongly coupled position–orientation operator, we solve the eigenvalue problem using a Galerkin spectral method and retain only the downstream-admissible modes. The non-self-adjointness renders the eigenfunctions non-orthogonal, so their coefficients are determined from the inlet flux through a weighted biorthogonal expansion. Excellent agreement between the theoretical predictions and individual-based simulations validates the theory. Applying the theory to plane Poiseuille flow in an advection-dominated regime, we characterize the downstream transport through vertical concentration profiles, two-dimensional concentration fields, cross-sectional position–orientation distributions and streamwise concentration distributions.

For passive particles, wall-normal redistribution is driven solely by translational diffusion, so the downstream development spans a relatively long streamwise scale and the vertical profile broadens monotonically from the point-source input towards uniformity. By contrast, even when the far-field profile remains uniform, as for spherical active particles, swimming produces a much richer and strongly non-monotonic downstream evolution. Starting from the downstream release orientation, rotational diffusion induces wall-normal swimming, after which local shear reorients the particles as they migrate outwards while being advected downstream. As they approach the upstream orientation, their wall-normal and streamwise velocities decrease, prolonging residence and producing two off-centre accumulation regions. For spherical particles, continued rotation redirects the population towards the centreline, producing split–return dynamics manifested as alternating off-centre and centreline accumulation regions. Dispersion progressively damps this redistribution, ultimately yielding a uniform far field. For ellipsoidal particles, orientation-dependent Jeffery rotation slows reorientation near both downstream and upstream alignment, preserving part of the centreline plume and prolonging residence near the lateral turning regions. This produces an intermediate three-peaked vertical profile and stronger, more persistent off-centre accumulation; farther downstream, the reduced coherence of centreward migration suppresses the secondary centreline accumulation, and the vertical distribution ultimately approaches a non-uniform far-field state.

The cross-sectional distributions reveal pronounced position–orientation coherence in the early developing region, with narrow curved ridges reflecting strong coupling between cross-stream migration and orientational dynamics. For spherical particles, orientational accumulation alternates between the downstream and upstream directions, tracing the coherent angular progression that generates successive off-centre and centreline accumulation regions. Ellipsoidal particles instead exhibit a more persistent, strongly non-uniform orientational structure, with downstream- and upstream-oriented populations coexisting over an extended region. Farther downstream, dispersion progressively weakens this coherence, and the orientational structure changes little while the vertical distribution still differs markedly from its fully developed form. Early downstream development occurs mainly in the centreline and two off-centre regions, producing alternating increases and decreases in the corresponding streamwise concentration profiles. Once this alternation ceases, appreciable spreading towards the walls begins, with ellipsoidal particles reaching the wall region earlier. Conservation of the total streamwise flux shows that the streamwise marginal concentration is the reciprocal of the concentration-weighted mean streamwise particle velocity and reflects the residence time per unit streamwise distance. In the fully developed far field, this concentration equals the reciprocal of the asymptotic drift velocity, establishing a relation between the continuous-release problem and the corresponding instantaneous-release dispersion problem. The non-uniform far-field distribution of ellipsoidal particles favours slower-moving states, leading to a higher streamwise marginal concentration than that of spherical particles.

The present study advances understanding of active-particle transport under continuous release, provides an important theoretical foundation for its analysis and points to several possible extensions. First, the current formulation neglects streamwise translational diffusion and treats the continuous-release problem in the downstream half-space. Extending it to the full space and retaining this diffusion would capture both the upstream motion of particles with negative streamwise velocities and diffusive spreading across the release plane \citep{novy_upstream_1990,huang_transport_2025}, thereby providing a more general description of the near-release region when particle transport into the upstream domain is non-negligible. Second, the present framework could be extended from steady continuous release to time-dependent release conditions, including periodically modulated sources, as previously considered in passive-solute dispersion \citep{gill_dispersion_1972,barton_dispersion_1983}. Third, the theory can be generalized beyond the two-dimensional channel geometry to more complex and higher-dimensional pipe-flow problems \citep{jiang_dispersion_2020}.

\appendix
\section{Individual-based numerical simulation for continuous release}
\label{app:individual_based_simulation}
The individual-based numerical simulations provide an independent
validation of the continuum solution. Each particle evolves according to
the dimensionless Langevin equations corresponding to the governing
equation \cref{eq dimensionless}. Introducing the dimensionless time
$t=D_r^{\ast}t^{\ast}$, the particle trajectories are advanced using the
Euler--Maruyama scheme. With streamwise translational diffusion neglected,
the provisional dimensionless update over one time step is
\begin{equation}
\left\{
\begin{aligned}
x_{n+1}
&=x_n+
\left[
U(y_n)+\frac{\mathit{Pe}_s}{\mathit{Pe}_f}\cos\theta_n
\right]\Delta t,\\
y_{n+1}
&=y_n+\mathit{Pe}_s\sin\theta_n\,\Delta t
+\sqrt{2D_t\Delta t}\,\xi^y_n,\\
\theta_{n+1}
&=\theta_n+
\frac{1}{2}\mathit{Pe}_f
\frac{\mathrm{d}U(y_n)}{\mathrm{d}y}
\left(-1+\alpha_0\cos2\theta_n\right)\Delta t
+\sqrt{2\Delta t}\,\xi^\theta_n .
\end{aligned}
\right.
\label{eq individual euler step}
\end{equation}
Here the subscript $n$ denotes the time-step index and $\Delta t$ is the
time-step size. The variables $\xi^y_n$ and $\xi^\theta_n$ are independent
standard normal random variables, sampled independently for each particle
at every time step.

For specular reflection at the channel walls, a provisional position outside
the channel is corrected according to
\begin{equation}
\left.
\begin{aligned}
y_{n+1}&\rightarrow 2-y_{n+1},
&\theta_{n+1}&\rightarrow-\theta_{n+1},
&&\text{if }y_{n+1}>1,\\
y_{n+1}&\rightarrow-y_{n+1},
&\theta_{n+1}&\rightarrow-\theta_{n+1},
&&\text{if }y_{n+1}<0.
\end{aligned}
\right\}
\label{eq specular reflection discrete}
\end{equation}
After the wall correction, $\theta_{n+1}$ is mapped periodically onto the
chosen $2\pi$-periodic orientation interval.

Direct simulation of continuous particle release would require new particles
to be introduced throughout the calculation, causing the number of tracked
particles and the computational cost to increase continually with time. To
avoid this, a fixed ensemble of particles is released instantaneously from
the prescribed source, and the steady continuous-release concentration field
is reconstructed by accumulating the resulting particle distributions.

All $N_p$ particles are initialized at $(x,y,\theta)=(0,1/2,0)$. Let
$\widehat{P}_{I,\ell}(x,y,\theta)$ denote the normalized ensemble
histogram of the instantaneous-release response at the $\ell$th
sampling instant. At each selected streamwise position, the histogram is constructed using particles within a narrow streamwise bin and is normalized by $N_p$ and the corresponding bin volume. For the unit dimensionless release rate considered here, the steady continuous-release concentration is approximated by
\begin{equation}
\widehat{P}(x,y,\theta)
=
\Delta t_s
\sum_{\ell=1}^{N_s}
\widehat{P}_{I,\ell}(x,y,\theta),
\label{eq:continuous_release_reconstruction}
\end{equation}
where $\Delta t_s$ is the sampling interval, chosen as an integer
multiple of $\Delta t$, and $N_s$ is the number of samples.

The simulations use $N_p=10^5$ particles and a time step
$\Delta t=10^{-4}$. The temporal accumulation is continued to
$N_s\Delta t_s=200$, which was found sufficient for convergence of the
reconstructed continuous-release concentration field. Since each sampled
histogram is weighted by $\Delta t_s$, the accumulated contribution of a
particle to each bin is proportional to the time it spends there. The vertical concentration profiles are obtained by integrating
$\widehat{P}$ over $\theta$, while the streamwise marginal concentration
is calculated directly from a one-dimensional histogram in $x$ using the
same temporal-accumulation procedure. This direct calculation is
equivalent to integrating $\widehat{P}$ over the cross-sectional space,
but avoids constructing a two-dimensional histogram at every streamwise
position.

\backsection[Funding]{This work is supported by the National Natural Science Foundation of China (grant no. 12372379).
}

\backsection[Declaration of interests]{The authors report no conflict of interest.}


\backsection[Author ORCID]{
	\urlstyle{same}
        \\
	Hanhan Zeng \url{https://orcid.org/0009-0003-7279-5906};
	\\
	Guoqian Chen \url{https://orcid.org/0000-0003-1173-6796}.}

\FloatBarrier
\bibliographystyle{jfm} 
\bibliography{mylibrary}

\end{document}